\documentclass[11pt,twoside]{article}
\usepackage{times,fancyhdr}
\usepackage{cite}
\usepackage{graphicx}
\usepackage{axodraw}

\def\beq{\begin{equation}}
\def\eeq{\end{equation}}
\def\beqa{\begin{eqnarray}}
\def\eeqa{\end{eqnarray}}

\date{}

\title{Resummations in QCD hard-scattering\\ at large and small $x$}
\author{Nikolaos Kidonakis$^1$, Agust{\' \i}n Sabio Vera$^2$, 
Philip Stephens$^1$\\ \\
\hspace{-4mm}$^1${\sl Kennesaw State University, 1000 Chastain Rd. \# 1202,
Kennesaw, GA 30144, USA}\\
\hspace{-4mm}$^2${\sl Physics Department, Theory Division, CERN, CH-1211 Geneva 23, Switzerland}}
 
\begin{document} 

\pagestyle{fancy}
\fancyhead{}
\fancyhead[EC]{Nikolaos Kidonakis, Agust{\' \i}n Sabio Vera, Philip Stephens}
\fancyhead[EL,OR]{\thepage}
\fancyhead[OC]{Resummations in QCD hard-scattering at large and small $x$}
\fancyfoot{} 
\renewcommand\headrulewidth{0.5pt}
\addtolength{\headheight}{2pt} 

\maketitle 

We discuss different resummations of large logarithms that arise in
hard-scattering cross sections of quarks and gluons in regions of large
and small $x$. The large-$x$ logarithms are typically dominant near threshold 
for the production of a specified final state. These soft and collinear gluon 
corrections produce large enhancements of the cross section for many processes,
notably top quark and Higgs production, and typically the higher-order 
corrections reduce the factorization and renormalization scale dependence 
of the cross 
section. The small-$x$ logarithms are dominant in the regime where the 
momentum transfer of the hard sub-process is much smaller than the total 
collision energy. These logarithms are important to describe multijet final 
states in deep inelastic scattering and hadron colliders, and in the study 
of parton distribution functions. The resummations 
at small and
large $x$ are linked by the eikonal approximation and are dominated by soft 
gluon anomalous dimensions. We 
will review their role in both contexts and provide some explicit calculations
at one and two loops.

\section{Introduction}

Particle physics in high-energy hadron colliders depends crucially on our 
ability to calculate cross sections to an ever increasing theoretical 
accuracy, which is achieved by the 
incorporation of higher-order corrections.
Hard-scattering cross sections in perturbative QCD obey factorization theorems
\cite{factor} that play a key role in the calculation of these corrections.
Typically, the cross section for a process involving the collision of two 
hadrons (proton-antiproton at the Fermilab Tevatron or proton-proton at 
the CERN LHC) into a specified final state can be described as a 
convolution of non-perturbative parton distribution functions that 
describe the parton content of the hadron, and a partonic cross section 
that can be calculated order-by-order in perturbation theory.
The short-distance partonic cross section involves the scattering of quarks 
and gluons.
The partonic processes are of the form
\beq
f_{1}(p_1)\, + \, f_{2}\, (p_2) \rightarrow F(p)\, + \, X \, ,
\eeq
where $f_1$ and $f_2$ represent partons (quarks or gluons),
$F$ represents an observed system in the final state, such as a top quark 
or a jet or a Higgs boson, and $X$ represents any additional 
final-state particles. 
The factorization is described schematically by 
\beqa
\sigma_{h_1 h_2 \rightarrow F}
=\sum_f \int  dx_1 dx_2 \, \phi_{f_1/h_1}(x_1,\mu_F)\,
\phi_{f_2/h_2}(x_2,\mu_F)\,
{\hat \sigma}_{f_1 f_2 \rightarrow F}(s,t,u,\mu_F,\mu_R) \, ,
\label{factcs}
\eeqa
where $\sigma_{h_1 h_2 \rightarrow F}$ is the physical cross section 
(total or differential) for the 
production of final state $F$ in the scattering of hadrons $h_1$ and $h_2$,
$\phi_{f_i/h_i}$ is the distribution function for parton $f_i$ with momentum 
fraction $x_i$ of hadron $h_i$, and ${\hat \sigma}_{f_1 f_2 \rightarrow F}$ 
is the partonic cross section. 
The collinear singularities are factorized in a process-independent
manner and absorbed into the parton 
distribution functions which are dependent on the factorization scale $\mu_F$.
The physical cross section is in principle independent of the 
factorization scale $\mu_F$ and the renormalization scale $\mu_R$, but in 
practice there is a strong dependence because we truncate the infinite 
perturbative series at finite order (typically next-to-leading-order (NLO) 
or next-to-next-to-leading-order (NNLO) in the strong coupling $\alpha_s$).
The parton-level cross section explicitly involves the standard kinematical
invariants,  $s=(p_1+p_2)^2$, $t=(p_1-p)^2$, $u=(p_2-p)^2$, 
formed from the 4-momenta of the particles in 
the hard scattering.

Near threshold, i.e. when the energy of the incoming partons is just 
sufficient to produce a final state without additional radiation, 
the production cross section receives significant corrections from large-$x$ 
logarithms \cite{GS,CT,KS,CLS}. These logarithms arise from incomplete 
cancellations between 
virtual terms and terms that describe soft-gluon emission. Since near 
threshold any additional radiation has to be soft, the large-$x$ logarithms 
are especially important in that kinematical region. Large-$x$ resummation 
depends critically on the color structure of the process 
\cite{KS,KOS1,KOS2,NKrev} as well as the kinematics \cite{KS,LOS}. 

Small-$x$ logarithms arise when the perturbative scales characterizing the 
hard-subprocess are much smaller than the
total collision energy. In this case resummation of logarithms of the
form $\ln(1/x)$ becomes important. When the transverse scales of the outgoing 
scattered particles are similar and large this resummation can be described
by the Balitsky-Fadin-Kuraev-Lipatov (BFKL) evolution equation 
\cite{BFKL1,BFKL2,BFKL3,BFKL4,BFKL5}. 
This equation is a linear integral equation which leads to an 
exponential rise of the cross section. The slope of this rise can be 
interpreted as a perturbative construction of the QCD Pomeron. This
Pomeron is considered the mediator of many QCD diffractive processes, such
as diffractive vector meson production. 

The conditions by which the BFKL evolution should be valid are satisfied by
jet production with large rapidity gaps. Phenomenological studies of this
process with the summation of the terms $\alpha_S^n \ln^n(1/x)$ (leading-order
kernel) are not very predictive since the value of the coupling is a free 
parameter and the Regge energy scale, a sort of factorization scale at high 
energies, can only be fixed at higher orders. Inclusion of the 
next-to-leading order corrections,
$\alpha_S^{n+1} \ln^n(1/x)$ \cite{Fadin:1998py,Ciafaloni:1998gs}, 
brings the predictions in closer 
agreement with data. 

In the next section we discuss large-$x$ resummation and finite-order 
expansions of the resummed cross section through 
next-to-next-to-next-to-leading order (NNNLO). 
In Section 3 we present some applications of large-$x$ resummation to 
various hard-scattering processes, namely top-antitop pair production, 
single top quark production, $W$-boson production at large transverse momentum,
and Higgs boson production via $b{\bar b} \rightarrow H$. 
In Section 4 we present typical one-loop and two-loop calculations 
in the eikonal approximation that are needed in resummations for processes 
with massive quarks, such as heavy quark pair production. Section 5 discusses 
small-$x$ resummation and applications of BFKL. We conclude in Section 6.

\section{Large-$x$ resummations}

Large-$x$ resummations depend crucially on the kinematics and 
color structure of the process under study.
In single-particle-inclusive (1PI) kinematics 
we identify one particle $F$ with momentum $p$.
In pair-invariant-mass (PIM) kinematics we identify a pair of particles
(such as a heavy quark-antiquark pair) with invariant mass squared $Q^2$. 

In general, the partonic cross section $\hat{\sigma}$ includes soft 
corrections in the form of plus distributions ${\cal D}_l(x_{th})$ 
with respect to a kinematical variable 
$x_{th}$ that measures distance from threshold, with $l \le 2n-1$ at $n$-th 
order in $\alpha_s$ beyond the leading order. 
In 1PI kinematics, 
$x_{th}$ is usually denoted as $s_4$
(or $s_2$) and is defined by $s_4=s+t+u-\sum m^2$, 
where the sum is over the squared masses 
of all particles in the process. At threshold, $s_4=0$. 
The plus distributions are then of the form 
\beq
{\cal D}_l(s_4)\equiv\left[\frac{\ln^l(s_4/M^2)}{s_4}\right]_+ \, ,
\eeq
where $M$ is a hard scale relevant to the process, for example
the mass $m$ of a heavy quark or the transverse momentum of a jet.
The plus distributions are defined through their integral with the  
parton distribution functions by 
\beqa
\int_0^{s_{4 \, max}} ds_4 \, \phi(s_4) \left[\frac{\ln^l(s_4/M^2)}
{s_4}\right]_{+} &\equiv&
\int_0^{s_{4\, max}} ds_4 \frac{\ln^l(s_4/M^2)}{s_4} [\phi(s_4) - \phi(0)]
\nonumber \\ &&
{}+\frac{1}{l+1} \ln^{l+1}\left(\frac{s_{4\, max}}{M^2}\right) \phi(0) \, .
\label{1piplus}
\eeqa
In PIM kinematics, $x_{th}$ is usually denoted as $1-x$ or $1-z$, 
with $z=Q^2/s \rightarrow 1$ at threshold.  Then the 
plus distributions are of the form
\beq
{\cal D}_l(z)\equiv\left[\frac{\ln^l(1-z)}{1-z}\right]_+
\eeq
defined by
\beqa
\int_{z_{min}}^1 dz \, \phi(z) \left[\frac{\ln^l(1-z)}{1-z}\right]_{+} &\equiv&
\int_{z_{min}}^1 dz \frac{\ln^l(1-z)}{1-z} [\phi(z) - \phi(1)]
\nonumber \\ && 
{}+\frac{1}{l+1} \ln^{l+1}(1-z_{min}) \phi(1) \, .
\label{pimplus}
\eeqa
The highest powers of these distributions in the $n$th-order corrections
are the leading logarithms (LL) with $l=2n-1$,
the second highest are the next-to-leading logarithms (NLL) with $l=2n-2$, 
etc. (note that the counting of logarithms 
is different in the exponent and in the fixed-order expansions). 
These logarithms can be resummed in 
principle to all orders in perturbation theory.

\subsection{Exponentiation}

The resummation of threshold logarithms is performed in moment space.
By taking moments, divergent distributions in $1-z$ (or $s_4$) produce 
powers of $\ln N$, with $N$ the moment variable:
\beq
\int_0^1 dz\; z^{N-1}\left[{\ln^m(1-z)\over 1-z}\right]_+
={(-1)^{m+1}\over m+1}\ln^{m+1}N +{\cal O}\left(\ln^{m-1}N\right)\, .
\eeq
If we define moments of the partonic cross section by
${\hat\sigma}(N)=\int dz \, z^{N-1} {\hat\sigma}(z)$ (PIM) or by  
${\hat\sigma}(N)=\int (ds_4/s) \;  e^{-N s_4/s} {\hat\sigma}(s_4)$ (1PI),
then the logarithms of $N$ that appear in ${\hat \sigma}(N)$ exponentiate.

The resummation follows from the factorization properties of the cross 
section. We begin the derivation of the resummed cross section by first 
writing a factorized form for the moment-space infrared-regularized 
parton-parton scattering cross section, 
$\sigma_{f_1 f_2 \rightarrow F}(N, \epsilon)$, 
which factorizes as the hadronic cross section
\beq
\sigma_{f_1 f_2 \rightarrow F}(N, \epsilon)  
={\tilde \phi}_{f_1/f_1}(N,\mu_F,\epsilon)\; 
{\tilde \phi}_{f_2/f_2}(N,\mu_F,\epsilon) \;
{\hat \sigma}_{f_1 f_2 \rightarrow F}(N,\mu_F,\mu_R) \, ,
\label{sigm}
\eeq
with the moments of $\phi$ given by  
$\tilde{\phi}(N)=\int_0^1dx\; x^{N-1}\phi(x)$.  
We factorize the initial-state collinear divergences, regularized by 
$\epsilon$, into the parton distribution functions, $\phi$, 
which are expanded to the 
same order in $\alpha_s$ as the partonic cross section, 
and we thus obtain the perturbative expansion for the
infrared-safe partonic short-distance function ${\hat \sigma}$.

The partonic short-distance function ${\hat \sigma}$ still has sensitivity
to soft-gluon dynamics through its $N$ dependence.   
We then refactorize the moments of the cross section
as \cite{KS,NKrev}
\beqa
&&\sigma_{f_1 f_2\rightarrow F}(N,\epsilon)
={\tilde\psi}_{f_1/f_1} \left(N,\mu_F,\epsilon \right) \;
{\tilde\psi}_{f_2/f_2} \left(N,\mu_F,\epsilon \right)  
\nonumber \\ && \hspace{-10mm} \times \; 
H_{IL}^{f_1 f_2\rightarrow F} \left(\alpha_s(\mu_R)\right)\; 
{\tilde S}_{LI}^{f_1 f_2 \rightarrow F} 
\left({M\over N \mu_F },\alpha_s(\mu_R) \right)\;
\prod_j  {\tilde J_j}\left (N,\mu_F,\epsilon \right)  
+{\cal O}(1/N) \, ,
\label{sigp}
\eeqa 
where $\psi$ are center-of-mass distributions that absorb 
the universal collinear singularities from the incoming partons, 
$H_{IL}$ are $N$-independent hard components  
which describe the hard-scattering, $S_{LI}$ is a soft gluon function
associated with non-collinear soft gluons, and $J$ are functions 
that absorb the collinear singularities from massless partons, if any, 
in the final state. 

$H$ and $S$ are matrices in color space and 
we sum over the color indices $I$ and $L$ 
that describe the color structure of the hard scattering.
The hard-scattering function involves contributions from the 
amplitude of the process and the complex conjugate of the amplitude,
$H_{IL}=h_L^*\, h_I$.
The soft function $S_{LI}$ represents the coupling of soft gluons to the
partons in the scattering. 
The color tensors of the hard scattering connect
together the eikonal lines to which soft gluons couple.
One can construct an eikonal operator describing 
soft-gluon emission and write a dimensionless eikonal cross section,
which describes the emission of soft gluons by the eikonal lines 
\cite{KS,KOS1,KOS2,NKrev}.

Comparing Eqs.\ (\ref{sigm}) and (\ref{sigp}), we
see that the moments of the short-distance partonic cross section are given by 
\beqa
{\hat \sigma}_{f_1 f_2\rightarrow F}(N,\mu_F,\mu_R)&=&
{{\tilde\psi}_{f_1/f_1}(N,\mu_F,\epsilon)\, 
{\tilde\psi}_{f_2/f_2}(N,\mu_F,\epsilon)
\over{\tilde \phi}_{f_1/f_1}(N,\mu_F,\epsilon)\, 
{\tilde \phi}_{f_2/f_2}(N,\mu_F,\epsilon)} \,
H_{IL}^{f_1 f_2\rightarrow F}\left(\alpha_s(\mu_R)\right)
\nonumber \\ &&  \times \;
{\tilde S}_{LI}^{f_1 f_2\rightarrow F}
\left(\frac{M}{N\mu_F},\alpha_s(\mu_R)\right) 
\prod_j  {\tilde J_j}\left (N,\mu_F,\epsilon \right) \, .
\label{psiphi}
\eeqa
All the factors in Eq.~(\ref{psiphi}) are gauge and factorization
scale dependent. The constraint that the product of these factors must be
independent of the gauge and factorization scale results
in the exponentiation of logarithms of $N$ in $\psi/\phi$ and $S_{LI}$ 
\cite{KS,CLS}. 

The soft matrix $S_{LI}$ depends on $N$ through the ratio $M/(N\mu_F)$,
and it requires renormalization as a composite operator.
Its $N$-dependence can thus be resummed by renormalization group
analysis \cite{KoRa,BoSt,GK,KK}.
However, the product $H_{IL}S_{LI}$ needs
no overall renormalization, because the UV divergences of $S_{LI}$
are balanced by those of $H_{IL}$.
Thus, we have \cite{KS,NKrev}
\beqa
H^{0}_{IL}&=& \prod_{i=a,b} Z_i^{-1}\; \left(Z_S^{-1}\right)_{IC}
H_{CD} \left[\left(Z_S^\dagger \right)^{-1}\right]_{DL} \, ,
\nonumber \\ 
S^{0}_{LI}&=&(Z_S^\dagger)_{LB}S_{BA}Z_{S,AI},
\label{HSren}
\eeqa
where $H^{0}$ and $S^{0}$ denote the unrenormalized quantities,
$Z_i$ is the renormalization constant of the $i$th
incoming partonic field, and $Z_S$ is
a matrix of renormalization constants, which describe the
renormalization of the soft function.  $Z_{S}$ is defined to include the
wave function renormalization
necessary for the outgoing eikonal lines that
represent any heavy quarks.

From Eq.\ (\ref{HSren}), we see that 
the soft function  $S_{LI}$ satisfies the
renormalization group equation~\cite{KS,KOS1,KOS2,NKrev}
\begin{equation}
\left(\mu {\partial \over \partial \mu}
+\beta(g_s){\partial \over \partial g_s}\right)\,S_{LI}
=-(\Gamma^\dagger_S)_{LB}S_{BI}-S_{LA}(\Gamma_S)_{AI}\, ,
\label{RGE}
\end{equation}
where $\beta$ is the QCD beta function and $g_s^2=4\pi\alpha_s$.
$\Gamma_S$ is an anomalous dimension matrix 
that is calculated in the eikonal approximation 
by explicit renormalization of the soft function.
In a minimal subtraction renormalization scheme and with
$\epsilon=4-n$, where $n$ is the number of space-time dimensions,
the soft anomalous dimension matrix is given at one loop by
\begin{equation}
\Gamma_S^{(1l)} (g_s)=-\frac{g_s}{2} \frac {\partial}{\partial g_s}
{\rm Res}_{\epsilon\rightarrow 0} Z_S (g_s, \epsilon) \, .
\label{GammaS1l}
\end{equation}
The process-dependent matrices $\Gamma_S$ have been 
calculated at one loop for all $2 \rightarrow 2$ 
partonic processes; a compilation of results is given in \cite{NKrev}. 
In processes with trivial or simple color structure $\Gamma_S$ is 
simply a function ($1\times 1$ matrix) while in processes with complex 
color structure it is a non-trivial matrix in color exchange.
For quark-(anti)quark scattering,  $\Gamma_S$ is a $2\times 2$ 
matrix \cite{KS,BoSt};
for quark-gluon scattering it is a $3\times 3$ matrix \cite{KOS2};  
for gluon-gluon scattering it is an $8\times 8$ matrix \cite{KOS2}.
Complete two-loop calculations of soft anomalous dimensions 
for processes with massless quarks have appeared 
in \cite{ADS}. Selected two-loop results for heavy quark production 
appeared in \cite{NK2l}.
We present a sample one-loop calculation in Section 4.1 
and a sample two-loop calculation in Section 4.2, both with outgoing 
massive quarks (see \cite{NKPS}).  

The exponentiation of logarithms of $N$ in the ratios $\psi/\phi$ and 
in the functions $J$ in Eq. (\ref{psiphi}), together with the solution of 
the renormalization group equation (\ref{RGE}), 
provide us with the complete expression for the resummed partonic cross 
section in moment space \cite{KS,KOS1,KOS2,NKrev,NKuni,NKNNNLO}
\beqa
{\hat{\sigma}}^{res}(N) &=&   
\exp\left[ \sum_i E^{f_i}(N_i)\right] \; 
\exp\left[ \sum_j {E'}^{f_j}(N_j)\right] 
\nonumber\\ && \hspace{-10mm} \times \,
\exp \left[\sum_i 2\int_{\mu_F}^{\sqrt{s}} \frac{d\mu}{\mu}\;
\gamma_{f_i/f_i}\left(\alpha_s(\mu)\right)\right] \;
\exp\left[2\, d_{\alpha_s} \int_{\mu_R}^{\sqrt s} \frac{d\mu}{\mu}\; 
\beta\left(\alpha_s(\mu)\right)\right] 
\nonumber\\ && \hspace{-10mm} \times \,
{\rm Tr} \left \{H^{f_1 f_2 \rightarrow F}\left(\alpha_s(\mu_R)\right) \;
\exp \left[\int_{\sqrt{s}}^{{\sqrt{s}}/{\tilde N_j}} 
\frac{d\mu}{\mu} \;
\Gamma_S^{\dagger \, f_1 f_2 \rightarrow F}\left(\alpha_s(\mu)\right)\right] 
\right. 
\nonumber\\ && \hspace{-10mm} \left. \times \,
{\tilde S^{f_1 f_2 \rightarrow F}} \left(\alpha_s\left(\frac{\sqrt{s}}
{\tilde N_j}\right) \right) 
\exp \left[\int_{\sqrt{s}}^{{\sqrt{s}}/{\tilde N_j}} 
\frac{d\mu}{\mu}\; \Gamma_S^{f_1 f_2 \rightarrow F}
\left(\alpha_s(\mu)\right)\right] \right\} \, .
\label{resHS}
\eeqa
The sums over $i=1,2$ run over incoming partons. The sum over $j$ is over 
massless partons, if any, in the final state at lowest order. 
The resummed expression is valid for either 1PI or PIM kinematics.
In 1PI kinematics $N_i=N (-t_i/M^2)$, where $t_i$ denotes $t$ or $u$, 
and  $N_j=N (s/M^2)$, while 
in PIM kinematics $N_i=N_j=N$. 
Also ${\tilde N}=N e^{\gamma_E}$, with $\gamma_E$ the Euler constant.

The first exponent in Eq. (\ref{resHS}) arises from the exponentiation
of logarithms of $N$ in the ratios $\psi/\phi$ of Eq. (\ref{psiphi}), 
and is given in the $\overline{\rm MS}$ scheme by 
\beq
E^{f_i}(N_i)=
-\int^1_0 dz \frac{z^{N_i-1}-1}{1-z}\;
\left \{\int^1_{(1-z)^2} \frac{d\lambda}{\lambda}
A_i\left(\alpha_s(\lambda s)\right)
+{\nu}_i\left[\alpha_s((1-z)^2 s)\right]\right\} \, ,
\label{Eexp}
\eeq
with $A_i(\alpha_s) = \sum_{n=1}^{\infty} (\alpha_s/\pi)^n A_i^{(n)}$.
At one loop, $A_i^{(1)}=C_i$ which is $C_F=(N_c^2-1)/(2N_c)$ for a quark 
or antiquark and $C_A=N_c$ for a gluon, with $N_c$ the number of colors,
while $A_i^{(2)}=C_i K/2$ with $K= C_A (67/18-\zeta_2) - 5n_f/9$ 
\cite{KoTr},
where $n_f$ is the number of quark flavors and $\zeta_2=\pi^2/6$. 
Also ${\nu}_i=\sum_{n=1}^{\infty}(\alpha_s/\pi)^n {\nu}_i^{(n)}$, 
with ${\nu}_i^{(1)}=C_i$. 

The second exponent in Eq. (\ref{resHS}) arises from the exponentiation of 
logarithms of $N$ in the functions $J_j$ of Eq. (\ref{psiphi}), 
and is given by 
\beqa
{E'}^{f_j}(N_j)&=&
\int^1_0 dz \frac{z^{N_j-1}-1}{1-z}\;
\left \{\int^{1-z}_{(1-z)^2} \frac{d\lambda}{\lambda}
A_j \left(\alpha_s\left(\lambda s\right)\right)
-B_j\left[\alpha_s((1-z)s)\right] \right.
\nonumber \\ && \hspace{30mm} \left. 
{}-\nu_{j}\left[\alpha_s((1-z)^2 s)\right]\right\} \, .
\label{Ejexp}
\eeqa
Here $B_j=\sum_{n=1}^{\infty} (\alpha_s/\pi)^n B_j^{(n)}$ 
with $B_j^{(1)}$ equal to $3C_F/4$ for quarks and $\beta_0/4$ for gluons, 
where $\beta_0=(11 C_A-2 n_f)/3$ is the lowest-order $\beta$ function.

The third exponent in Eq. (\ref{resHS}) 
controls the factorization scale dependence of the cross 
section, and $\gamma_{f_i/f_i}$ is the moment-space 
anomalous dimension of the ${\overline {\rm MS}}$ density $\phi_{f_i/f_i}$.
The $\beta$ function in the fourth exponent controls the renormalization 
scale dependence of the cross section.
The constant $d_{\alpha_s}$ takes the value $k$ if the Born cross 
section is of order $\alpha_s^k$.
Explicit expressions for the functions in these four exponents, and related 
references, are assembled for convenience in Appendix A of Ref. \cite{NKHiggs}.

As noted before, both $H$ and $S$ are process-dependent matrices in 
color space and thus the trace is taken in Eq. (\ref{resHS}). 
At lowest order, the trace of the product of $H$ and $S$
reproduces the Born cross section.
The evolution of the soft function $S$  
follows from its renormalization group equation, (\ref{RGE}), and 
is given in terms of the soft anomalous dimension matrix $\Gamma_S$.

\subsection{NNNLO expansions}

The exponentials in the resummed partonic cross section can be expanded 
to any fixed order in $\alpha_s$ and then inverted to momentum space
to provide explicit results for the higher-order corrections.
A fixed-order expansion avoids using a prescription to regulate the 
infrared singularities 
in the exponents and thus no prescription is needed to deal with these 
in this approach (see discussion in Ref. \cite{NKtop}).

We now expand the resummed cross section, Eq. (\ref{resHS}), in 1PI kinematics 
through NNNLO. We provide results 
here for the case where $\Gamma_S$ are trivial (1 $\times$ 1) color matrices.
Explicit expressions for the more general case are found through NNLO 
in \cite{NKuni} and through NNNLO in \cite{NKNNNLO}.

At NLO, we find the expression for the soft-gluon corrections 
\beq
{\hat{\sigma}}^{(1)} = \sigma^B \frac{\alpha_s(\mu_R^2)}{\pi}
\left\{c_3\, {\cal D}_1(s_4) + c_2\,  {\cal D}_0(s_4) 
+c_1\,  \delta(s_4)\right\}
\label{NLOmaster}
\eeq
where $\sigma^B$ is the leading-order (LO) term, the LL coefficient is 
\beq
c_3=\sum_i 2 \, C_i -\sum_j C_j\, ,
\label{c3}
\eeq 
with $C_q=C_F$ and $C_g=C_A$,
and the NLL coefficient $c_2$ is defined by $c_2=c_2^{\mu}+T_2$, 
with
\beq
c_2^{\mu}=-\sum_i C_i \ln\left(\frac{\mu_F^2}{M^2}\right)
\eeq
denoting the terms involving logarithms of the factorization scale, and  
\beqa
T_2&=&2 {\rm Re} \Gamma_S^{(1)} - \sum_i \left[C_i
+2 \, C_i \, \ln\left(\frac{-t_i}{M^2}\right)+
C_i \ln\left(\frac{M^2}{s}\right)\right]
\nonumber \\ &&
{}-\sum_j \left[B_j^{(1)}+C_j
+C_j \, \ln\left(\frac{M^2}{s}\right)\right] \, 
\label{c2n}
\eeqa
denoting the scale-independent terms.
Again, $t_i$ denotes $t$ or $u$, the sums over $i$ run over 
incoming partons, and the sums over $j$
run over any massless partons in the final state at LO. 

We write the NLO $\delta(s_4)$ terms as 
$c_1 =c_1^{\mu} +T_1$, where
\beq
c_1^{\mu}=\sum_i \left[C_i\, \ln\left(\frac{-t_i}{M^2}\right) 
-\gamma_i^{(1)}\right]\ln\left(\frac{\mu_F^2}{M^2}\right)
+d_{\alpha_s} \frac{\beta_0}{4} \ln\left(\frac{\mu_R^2}{M^2}\right) 
\label{c1mu}
\eeq
denotes the terms involving logarithms of the factorization and 
renormalization scales. Here $\gamma_q^{(1)}=3C_F/4$ and 
$\gamma_g^{(1)}=\beta_0/4$, and 
$T_1$ denotes virtual terms that cannot be derived from 
the resummation formalism but can be determined 
by matching to a full NLO calculation for any specified process.

At NNLO, the soft-gluon corrections are 
\beqa
{\hat{\sigma}}^{(2)}&=& 
\sigma^B \frac{\alpha_s^2(\mu_R^2)}{\pi^2} \;
\left\{\frac{1}{2} \, c_3^2 \; {\cal D}_3(s_4) 
+\left[\frac{3}{2} \, c_3 \, c_2 - \frac{\beta_0}{4} \, c_3
+\sum_j C_j \, \frac{\beta_0}{8}\right] \; {\cal D}_2(s_4) \right.
\nonumber \\ && \hspace{-15mm} \left.
{}+\left[c_3 \, c_1 +c_2^2
-\zeta_2 \, c_3^2 -\frac{\beta_0}{2} \, T_2 
+\frac{\beta_0}{4} \, c_3 \, \ln\left(\frac{\mu_R^2}{M^2}\right)
+c_3\, \frac{K}{2} -\sum_j\frac{\beta_0}{4} \, B_j^{(1)} \right] \;  
{\cal D}_1(s_4) \right\} 
\nonumber \\ &&
{}+\cdots 
\label{NNLOmaster}
\eeqa
where we show explicitly results through next-to-next-to-leading logarithms
(NNLL). For a complete expression see \cite{NKuni, NKNNNLO}.

At NNNLO, the soft-gluon corrections are 
\beqa
{\hat{\sigma}}^{(3)}&=& 
\sigma^B \frac{\alpha_s^3(\mu_R^2)}{\pi^3} \; \left\{ 
\frac{1}{8} \, c_3^3 \; {\cal D}_5(s_4)
+\left[\frac{5}{8} \, c_3^2 \, c_2 -\frac{5}{24} \, \beta_0 \, c_3^2 
+\frac{5}{48} \, c_3 \, \beta_0 \sum_j C_j \right] \;  
{\cal D}_4(s_4) \right.
\nonumber \\ && \hspace{-10mm}
{}+\left[c_3 \, c_2^2 +\frac{1}{2} \, c_3^2 \, c_1
-\zeta_2 \, c_3^3 +\frac{\beta_0^2}{12} c_3-\frac{\beta_0}{3} c_3 c_2
-\frac{\beta_0}{2} c_3 T_2 
+\frac{\beta_0}{4} c_3^2 \ln\left(\frac{\mu_R^2}{M^2}\right) \right.
\nonumber \\ && \hspace{-5mm} \left. \left.
{}+c_3^2 \frac{K}{2}
+c_2 \frac{\beta_0}{6} \sum_j C_j -c_3 \sum_j \frac{\beta_0}{4} B_j^{(1)}    
-\sum_j C_j \, \frac{3\beta_0^2}{48}\right] \; {\cal D}_3(s_4) \right\} +\cdots
\label{NNNLOmaster}
\eeqa
where again we show explicitly results through NNLL. 
The complete expression is given in \cite{NKNNNLO}.

In PIM kinematics we simply replace $s_4$ by $1-z$, 
set $s=M^2$, and drop the terms with  
$\ln(-t_i/M^2)$ in the above formulas.

The NNNLO master equation, (\ref{NNNLOmaster}), 
gives the structure of the soft corrections and can provide 
the full soft corrections explicitly if all the necessary 
two-loop and three-loop quantities are known. For processes 
with non-trivial color structure we are currently limited to NLL or 
NNLL accuracy. For processes with trivial color structure, such as 
$b{\bar b} \rightarrow H$ \cite{Ravi,NKHiggs}, all soft-gluon corrections 
have been determined through NNNLO.
Below, the term ``N$^{(n)}$LO-N$^{(l)}$LL'' means that the soft-gluon 
contributions through N$^{(l)}$LL accuracy to the $n$-th order QCD corrections 
have been included.

\section{Applications of large-$x$ resummations}

In this section we provide some calculations that are applications 
of the large-$x$ resummation formalism to processes of interest at the 
Tevatron and the LHC. We present results for top-antitop pair production, 
single top quark production, $W$-boson production, and Higgs boson production 
via $b{\bar b} \rightarrow H$.

\subsection{$t{\bar t}$ production}

The top quark, the heaviest known elementary particle, was discovered in 
$p {\bar p}$ collisions at Run I of the Tevatron in 1995 \cite{CDFtt,D0tt}. 
More recent measurements at Run II have increased the accuracy of the top mass
and cross section measurements (for a review see \cite{WW,Kehoe}) 
and thus require accurate 
theoretical calculations of top production cross sections and differential 
distributions. The main partonic channels in $t{\bar t}$ production are 
$q{\bar q} \rightarrow t{\bar t}$, which is dominant at the Tevatron, 
and $gg \rightarrow t{\bar t}$, which will be dominant at the LHC.

The latest calculation for top-antitop pair hadroproduction
includes NNLO soft-gluon corrections  
to the double differential cross section \cite{NKRVtop}.
Near threshold the soft-gluon corrections dominate the cross section 
at the Tevatron and contribute sizable enhancements. 
The form of the corrections and their numerical values
depend crucially on the kinematics chosen to describe the process.
The NNLO soft corrections were calculated fully to 
NNLL in both 1PI and PIM kinematics \cite{NKtop,KLMV}. 
In addition a good approximation 
for the next-to-next-to-next-to-leading logarithms (NNNLL) was provided 
in \cite{NKRVtop}.  The best theoretical 
result for the cross section is the average of  
the NNLO-NNNLL cross sections in the two different kinematics \cite{NKRVtop}.

\begin{figure}
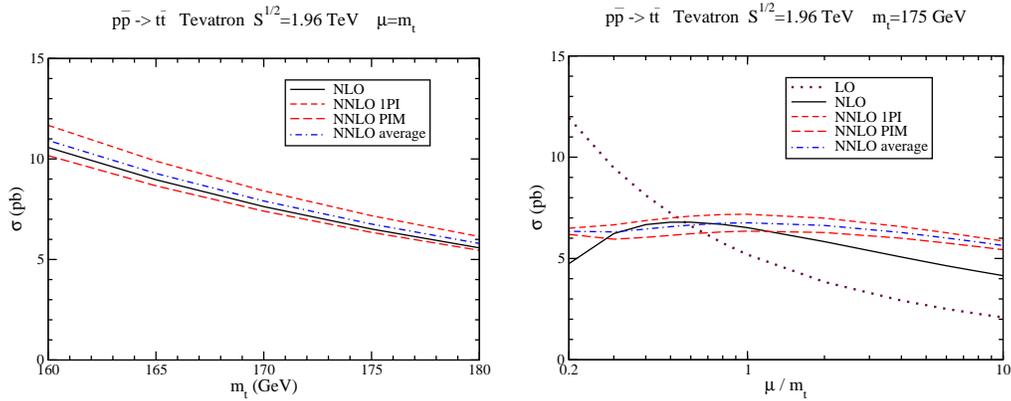

\centering
\includegraphics[width=0.46\textwidth]{ttbarmtplotnova.eps}
\hspace{3mm}
\includegraphics[width=0.46\textwidth]{ttbarmuplotnova.eps}
\caption[]{The $t \overline t$ total cross sections in $p \overline p$
collisions at the Tevatron with $\sqrt{S} = 1.96$ TeV  
as functions of $m_t$ (left) and $\mu/m_t$ (right).}
\label{ttbar}
\end{figure}

In Fig.~\ref{ttbar}, we present the NLO and approximate NNLO-NNNLL
$t \overline t$ cross sections at the Tevatron with $\sqrt{S}=1.96$ TeV 
using the MRST2002 \cite{MRST2002} parton densities. 
On the left we plot the cross sections as functions of $m_t$, 
the top quark mass, for $\mu = m_t$, where $\mu$ denotes the factorization 
and renormalization scales which we have set equal to each other. 
On the right we plot the cross sections as functions of 
$\mu/m_t$ with $m_t=175$ GeV.  
The results are given in both 1PI and PIM kinematics together with 
their average.
The NLO cross section depends less on $\mu$ than the LO cross 
section, as expected.  The NNLO-NNNLL cross sections exhibit even less 
dependence on $\mu$, approaching the scale independence of a true physical
cross section. They change by less than 3\% in the range 
$m_t/2 < \mu < 2m_t$.
For a top mass of 175 GeV the average of the NNLO-NNNLL 1PI and PIM results 
is $6.77 \pm 0.42$ pb, where the uncertainty indicated is from the kinematics.
Including all sources of uncertainty (kinematics, scale variation, and 
uncertainty from the parton distribution functions) we may write the 
cross section as $6.8 \pm 0.6$ pb. This theoretical result is in agreement 
with the latest experimental result for the cross section at the 
Tevatron \cite{CDFcs,D0cs}. Finally, we note that NNNLO soft-gluon 
contributions in the $q{\bar q} \rightarrow t{\bar t}$ channel were 
presented in \cite{NKNNNLO}. These NNNLO-NNLL corrections further 
stabilize the scale dependence of the cross section at the Tevatron. 

\subsection{Single top quark production}

Single top quark production provides a way to directly measure 
electroweak properties of the top quark, such as 
the $V_{tb}$ CKM matrix element. It also allows a deeper study
of electroweak theory since the top quark mass is of the same order of 
magnitude as the electroweak symmetry breaking scale, and may be useful 
in the discovery of new physics. Therefore it is crucial to have accurate
theoretical predictions for the cross section.

The cross section for single top quark production 
is less than the $t{\bar t}$ cross
section and the backgrounds to the production processes make the extraction 
of the single top signal challenging.
Intensive searches for single top quark events at the Tevatron 
have recently produced evidence of such events \cite{D0t,CDFt}.
The LHC has good potential for observation and further analysis 
of single top events.

Single top quarks can be produced through three distinct
partonic processes. One is $t$-channel production, $qb \rightarrow q' t$
and ${\bar q} b \rightarrow {\bar q}' t$, via
the exchange of a space-like $W$ boson, a second is $s$-channel
production, $q{\bar q}' \rightarrow {\bar b} t$, via the exchange of a 
time-like $W$ boson,
and a third is associated $tW$ production, $bg \rightarrow tW^-$.

The threshold corrections to single top production have been calculated 
for both the Tevatron and LHC colliders through NNNLO 
\cite{NKsntoptev,NKsntoplhc,NKsnh}.
At the Tevatron the $t$-channel process is numerically dominant, but the 
higher-order corrections are relatively small. The $s$-channel is 
smaller, but receives large corrections and it was shown that the threshold 
soft-gluon corrections dominate the cross section. 
Associated $tW$ production is quite minor, although it also has large $K$ 
factors, defined as the ratios of the higher-order cross sections to the LO 
cross section. 
At the LHC the $t$ channel is again dominant, but the second largest
channel is $tW$ production; the $s$ channel is numerically the smallest.
Below we provide some numerical results for all three channels at both the 
Tevatron and the LHC colliders using the MRST2004 parton densities 
\cite{MRST2004}. We add the soft-gluon corrections through NNNLO 
to the complete NLO cross section \cite{bwhl,Zhu}.

We begin with single top production at the Tevatron 
\cite{NKsntoptev} with $\sqrt{S}=1.96$ TeV. 
For $t$-channel production, the NNNLO-NLL cross section is 
$\sigma^{t-{\rm channel}}(m_t=175 \,{\rm GeV})=1.08^{+0.02}_{-0.01}\pm 0.06$
pb, where the first uncertainty is from variation of the factorization
and renormalization scales, $\mu_F$ and $\mu_R$, between $m_t/2$ and
$2m_t$, and the second is due to the parton distribution functions. 
For the $s$ channel, the corresponding cross section is
$\sigma^{s-{\rm channel}}(m_t=175 \,{\rm GeV})=0.49 \pm 0.02 \pm 0.01$ pb. 
Finally, in the $tW$ channel 
$\sigma^{tW}(m_t=175 \,{\rm GeV})=0.13 \pm 0.02 \pm 0.02$ pb.
We note that the cross sections for antitop production at the Tevatron 
are identical to those for single top production in each channel.

\begin{figure}
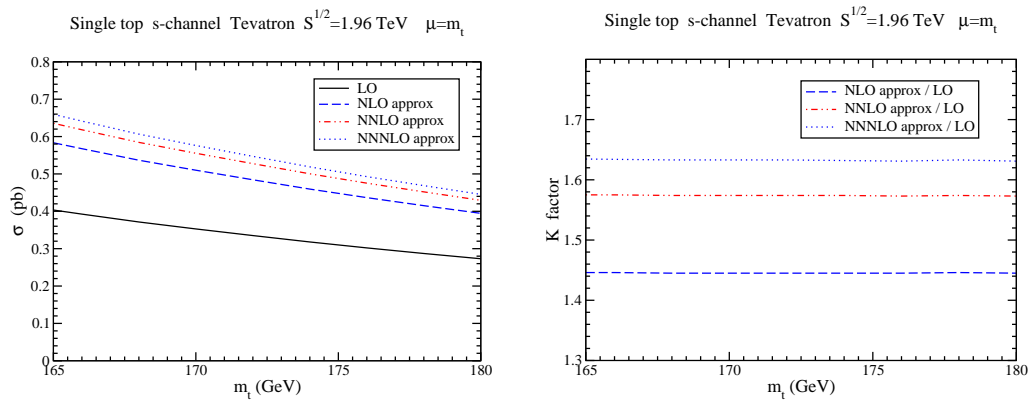

\centering
\includegraphics[width=0.46\textwidth]{schtevmtplotnova.eps}
\hspace{5mm}
\includegraphics[width=0.46\textwidth]{Kschtevmtplotnova.eps}
\caption[]{The cross section (left) and $K$ factors (right) for single top 
quark production at the Tevatron
in the $s$ channel. Here $\mu=\mu_F=\mu_R=m_t$.}
\label{csKschtevmtplot}
\end{figure}

In Fig. \ref{csKschtevmtplot} we plot the cross section and the $K$ factors   
for single top quark production at the Tevatron 
in the $s$ channel setting
both the factorization and renormalization scales to $\mu=m_t$.
We plot the LO cross section
and the approximate NLO, NNLO, and NNNLO cross sections at NLL accuracy.
The $K$ factors are quite large, thus showing that the corrections provide a
big enhancement to the cross section.
 
\begin{figure}
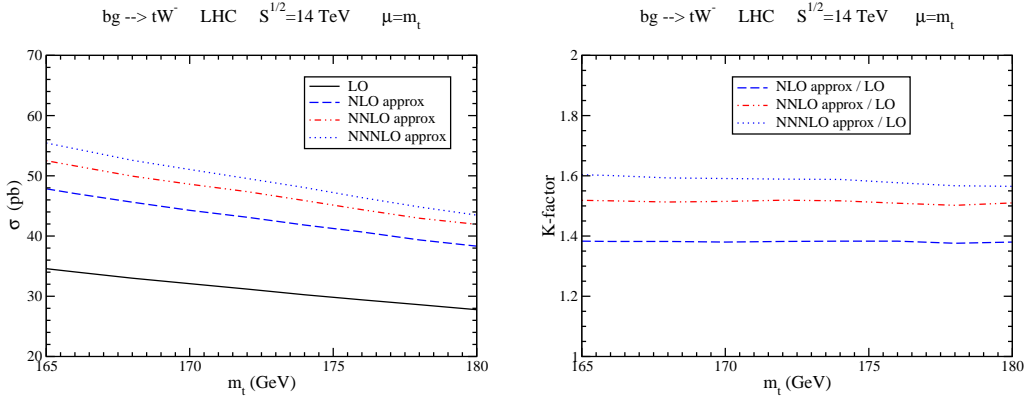

\centering
\includegraphics[width=0.46\textwidth]{bglhcmtplotnova.eps}
\hspace{5mm}
\includegraphics[width=0.46\textwidth]{Kbglhcmtplotnova.eps}
\caption[]{The cross section (left) and $K$ factors (right) 
for associated $tW$ production at the
LHC. Here $\mu=\mu_F=\mu_R=m_t$.}
\label{bglhcmtplot}
\end{figure}

We continue with single top production at the LHC \cite{NKsntoplhc} 
with $\sqrt{S}=14$ TeV.
For the $t$ channel the threshold corrections are not a good approximation 
of the complete corrections. The NLO cross section for top production
$\sigma^{t-{\rm channel}}_{\rm top} (m_t=175 \,{\rm GeV})=146 \pm 4 \pm 3$ pb.
For antitop production the corresponding result is 
$\sigma^{t-{\rm channel}}_{\rm antitop} (m_t=175 \,{\rm GeV})=89 \pm 3 \pm 2$ 
pb. 
For the $s$ channel,  the soft-gluon corrections are relatively large and 
the soft-gluon approximation is good. 
The NNNLO-NLL cross section is $\sigma^{s-{\rm channel}}_{\rm top}(m_t=175 \,{\rm GeV})
=7.23^{+0.53}_{-0.45}\pm 0.13$ pb for single top production and 
$\sigma^{s-{\rm channel}}_{\rm antitop}(m_t=175 \,{\rm GeV})
=4.03^{+0.10}_{-0.12}\pm 0.10$ pb for single antitop production.
Finally, for $tW$ production the cross section is 
$\sigma^{tW}(m_t=175 \,{\rm GeV})=41.1 \pm 4.1 \pm 1.0$ pb, which is identical 
to that for associated antitop production.
In Fig. \ref{bglhcmtplot} we plot the cross section and $K$ factors 
for associated $tW$ production at the LHC 
setting the scales to $\mu=m_t$. As seen from the plots, the soft-gluon 
corrections are large for this process.

\subsection{$W$-boson production at large transverse momentum}

$W$-boson production in hadron colliders can be used in 
testing the Standard Model and in estimating backgrounds to
Higgs production and new physics.
Precise calculations for $W$ production at large
transverse momentum, $Q_T$, are needed to identify 
signals of new physics which may be expected to
enhance the $Q_T$ distribution at high $Q_T$.

Analytical NLO calculations of the cross section for $W$ production at large 
transverse momentum were presented in Refs. \cite{AR,gpw}, where numerical 
results were also presented for the Fermilab Tevatron.
Numerical NLO results for $W$ production at the LHC were more recently 
presented in \cite{GKSV}.
The NLO corrections enhance the $Q_T$ distribution of the $W$ boson 
and they reduce the factorization and renormalization
scale dependence of the cross section.

A recent theoretical study \cite{NKASV} included soft-gluon corrections
through NNLO, which provide additional enhancements and a further reduction
of the scale dependence. The complete NNLL terms were calculated and an 
approximation for the NNNLL terms was derived at NNLO. 
Numerical results with these soft corrections 
were calculated for $W$ production at the Tevatron \cite{NKASV} and the 
LHC \cite{GKSV}.
 
Here we discuss $W$ production at large transverse momentum
at the LHC with $\sqrt{S}=14$ TeV using the MRST2002 
parton densities \cite{MRST2002}.
The LO partonic processes for the 
production of a $W$ boson and a parton are
$qg \rightarrow Wq$ and $q{\bar q} \rightarrow Wg$.
The electroweak coupling $\alpha(M_Z^2)$ is evaluated at the mass of the
$Z$ boson, and standard values \cite{pdg} are used for the various electroweak
parameters.
In the numerical results we present the sum of cross sections for
$W^-$ and $W^+$ production. 
The $W$ bosons at the LHC will be detected primarily through their leptonic
decay products e.g., $W^-\rightarrow \ell\bar\nu_\ell$, therefore 
the cross sections presented here should be multiplied by the appropriate
branching ratios.

\begin{figure}
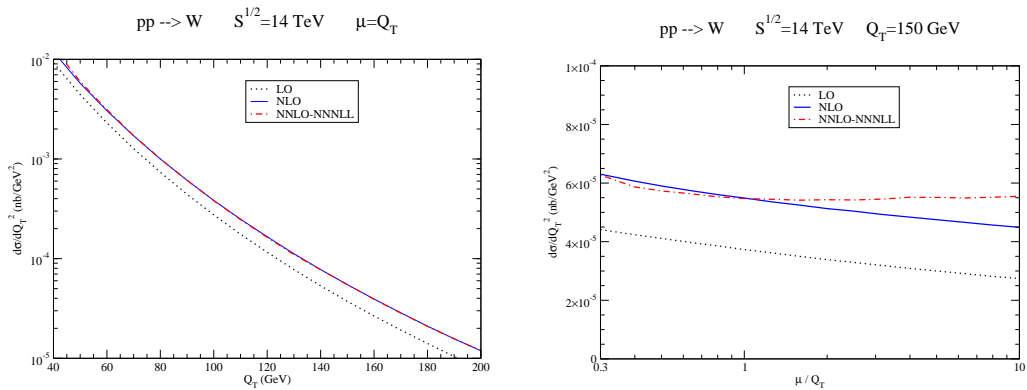

\vspace{6mm}
\centering
\includegraphics[width=0.46\textwidth]{wlhcplotnova.eps}
\hspace{5mm}
\includegraphics[width=0.46\textwidth]{mu150wlhcplotnova.eps}
\caption[]{The differential cross section,
$d\sigma/dQ_T^2$, for $W$ production  
at the LHC with $\mu=Q_T$ (left) and $Q_T=150$ GeV (right).}
\label{WLHC} 
\end{figure}

In Fig. \ref{WLHC} (left plot) we plot the transverse momentum 
distribution,
$d\sigma/dQ_T^2$, at high $Q_T$ for $W$ production at the LHC. 
We set $\mu_F=\mu_R=Q_T$ and denote this common
scale by $\mu$.
We plot LO, NLO, and NNLO-NNNLL results using the corresponding 
parton densities. 
As seen from the plot, the NLO corrections provide a significant enhancement 
of the LO $Q_T$ distribution. The NNLO-NNNLL corrections provide a
rather small further enhancement of the $Q_T$ distribution.
However, the NNLO-NNNLL corrections can be much bigger for
other choices of factorization and renormalization scales.
The NLO corrections
increase the LO result by about 30\% to 50\% in the $Q_T$ range shown.
In contrast, the NNLO-NNNLL/NLO ratio for this scale is rather small.
Part of the reason for this is that the NNLO parton distribution
functions are significantly smaller than the NLO pdf. 

On the plot on the right in Fig. \ref{WLHC} 
we show the scale dependence of $d\sigma/dQ_T^2$ 
for $Q_T=150$ GeV versus $\mu/Q_T$ over two
orders of magnitude. It is interesting to note that the scale dependence
of the cross section is not reduced when the NLO corrections are
included.
This is due to the fact that the cross section is dominated by
the process $qg\rightarrow Wq$.
The gluon density in the proton, at fixed $x$
less than ${\sim}0.01$, increases rapidly with scale.
Thus, the $\mu_R$ and $\mu_F$ dependencies cancel one another
to a large extent.
However, we have an improvement in the scale variation when the NNLO-NNNLL 
corrections are added. The NNLO-NNNLL result displays very little scale 
dependence.

\subsection{Higgs boson production via $b{\bar b} \rightarrow H$}

The search for the Higgs boson \cite{Higgs} is one of the most important goals
at the Tevatron and the LHC colliders \cite{HiggsWG}. The main Standard Model  
production channel at these colliders is $gg \rightarrow H$. However,
the channel $b{\bar b}\rightarrow H$ can be competitive 
in the Minimal Supersymmetric Standard Model at high $\tan\beta$, 
with $\tan\beta$ the ratio of the vacuum expectation  
values for the two Higgs doublets. The complete NNLO QCD corrections 
for this process were calculated in \cite{HKbb}. 

Complete expressions for the soft-gluon corrections at NNNLO were presented 
in \cite{NKHiggs,Ravi}. However, it is known at NNLO that the soft corrections 
alone are not a good approximation of the full corrections \cite{HKbb,NKHiggs}.
Purely collinear terms \cite{KLS,NKcoll,NKHiggs} have to be included to 
provide an accurate calculation. An approximation for the collinear terms 
through NNLL accuracy at NNNLO was provided in \cite{NKHiggs}.

\begin{figure}
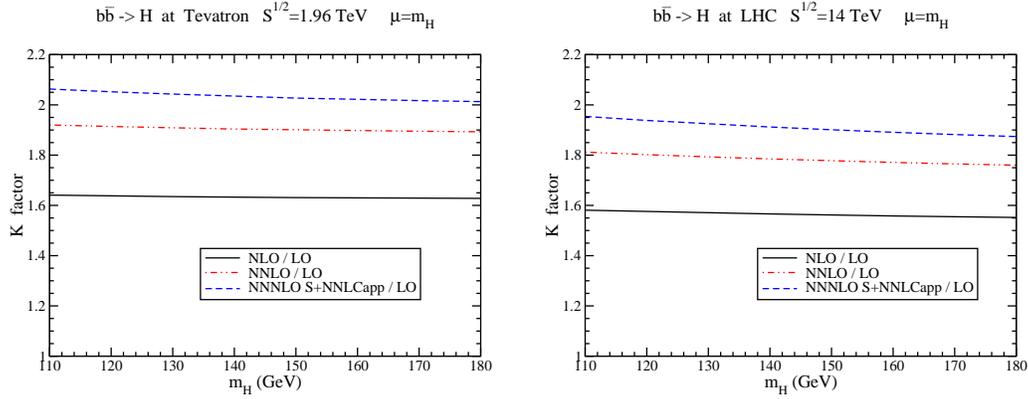

\centering
\includegraphics[width=0.46\textwidth]{Kbbhtevplotnova.eps}
\hspace{5mm}
\includegraphics[width=0.46\textwidth]{Kbbhlhcplotnova.eps}
\caption[]{The $K$ factors for $b {\bar b} \rightarrow H$ at the 
Tevatron (left) and the LHC (right). Here $\mu=\mu_F=\mu_R=m_H$.}
\label{Kbbhplot}
\end{figure}

We now present numerical results for $b{\bar b}\rightarrow H$ at the Tevatron 
and the LHC \cite{NKHiggs} using the MRST2006 parton densities \cite{MRST2006}.
Figure \ref{Kbbhplot} shows the $K$ factors for Higgs production via
$b {\bar b} \rightarrow H$ at the Tevatron (left)
and the LHC (right), with $\mu=m_H$.
The complete NLO corrections increase the LO 
result by around 60\% at both the Tevatron and the LHC. 
Inclusion of the complete NNLO corrections futher increases the cross section: 
the NNLO $K$ factor is around 1.9 at the Tevatron and
1.8 at the LHC. 
By including at NNNLO the sum of the complete soft-gluon 
corrections and the collinear approximate NNLL corrections
(S+NNLLapp), we find further enhancement. 
From the study of the contributions of the soft and collinear terms at NLO 
and NNLO at both the Tevatron and the LHC we expect that the NNNLO 
S+NNLCapp curve provides a good approximation of the complete NNNLO cross 
section.
The NNNLO S+NNLCapp $K$ factor is between 2.06 and 2.01 at the
Tevatron and between 1.95 and 1.87 at the LHC for Higgs masses ranging between
110 and 180 GeV, which is a significant addition to the NNLO result.
 
\section{Loop calculations in the eikonal approximation}

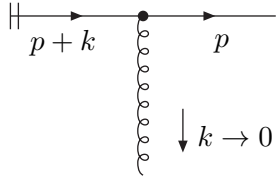
\begin{figure}[htb]
\begin{center}
\begin{picture}(120,120)(0,0)
\Line(0,80)(0,90)
\Line(3,80)(3,90)
\ArrowLine(0,85)(50,85)
\ArrowLine(50,85)(100,85)
\Vertex(50,85){2}
\Gluon(50,25)(50,85){2}{8}
\Text(20,75)[c]{$p+k$}
\Text(80,75)[c]{$p$}
\LongArrow(65,50)(65,35)
\Text(85,40)[c]{$ k \rightarrow 0$}
\end{picture}
\end{center}
\vspace{-10mm}
\caption{\label{eikonal}  Eikonal approximation.}
\end{figure}

The soft-gluon resummation formalism, and in particular the calculation 
of the soft anomalous dimension matrices, employs the use of the 
eikonal approximation in loop diagrams.
The eikonal approximation is valid for descibing the emission
of soft gluons from partons in the hard scattering. 
The approximation leads to a simplified form of the Feynman rules by 
removing the Dirac matrices from the calculation.
When the gluon momentum goes to zero,
the Feynman rules for the quark propagator and quark-gluon
vertex in Figure \ref{eikonal} simplify as follows:
\beq
{\bar u}(p) \, (-i g_s T_F^c) \, \gamma^{\mu}
\frac{i (p\!\!/+k\!\!/+m)}{(p+k)^2
-m^2+i\epsilon} \rightarrow {\bar u}(p)\,  g_s T_F^c \, \gamma^{\mu}
\frac{p\!\!/+m}{2p\cdot k+i\epsilon}
={\bar u}(p)\, g_s T_F^c \,
\frac{v^{\mu}}{v\cdot k+i\epsilon}
\eeq
with $v$ a dimensionless vector, $p \propto v$,  
and $T_F^c$ the generators of SU(3) in the fundamental representation.

The ultraviolet poles in loop diagrams involving eikonal lines 
are particularly important as they play a direct role in the renormalization
group evolution equations that are used in threshold resummations 
\cite{KS,KOS1,KOS2} (see Eq. (\ref{GammaS1l})).

Below we give examples of a one-loop and a two-loop calculation for 
diagrams involving eikonal lines representing massive quarks.
For the calculation we use the Feynman gauge, and we use dimensional 
regularization with $n=4-\epsilon$ dimensions.

\subsection{One-loop calculation}

\begin{figure}[htb]
\begin{center}
\begin{picture}(120,120)(0,0)
\Vertex(0,50){5}
\ArrowLine(0,50)(60,80)
\ArrowLine(60,80)(100,100)
\Vertex(60,80){2}
\Gluon(60,80)(60,20){2}{8}
\Text(25,78)[c]{$p_i+k$}
\Text(88,100)[c]{$p_i$}
\LongArrow(70,55)(70,45)
\Text(80,50)[c]{$k$}
\ArrowLine(0,50)(60,20)
\ArrowLine(60,20)(100,0)
\Vertex(60,20){2}
\Text(25,20)[c]{$p_j-k$}
\Text(88,0)[c]{$p_j$}
\end{picture}
\end{center}
\caption{\label{oneloop}  One-loop eikonal diagram with outgoing massive 
quarks.}
\end{figure}
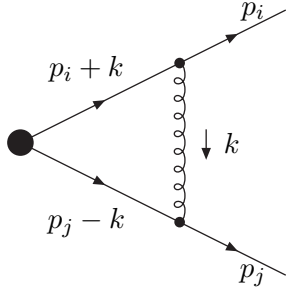

In this subsection we calculate the integral $I_{1l}$ for the one-loop 
diagram in Fig. (\ref{oneloop}) with eikonal lines representing outgoing 
massive quarks. This one-loop integral is given by
\beq
I_{1l} = g_s^2 \int\frac{d^n k}{(2\pi)^n} \frac{(-i)g_{\mu \nu}}{k^2} 
\frac{v_i^{\mu}}{v_i\cdot k} \, \frac{(-v_j^{\nu})}{(-v_j\cdot k)} \, .
\eeq
Using Feynman parameterization, this integral can be rewritten as
\beq
I_{1l} =-2i g_s^2 \, \frac{v_i \cdot v_j}{(2\pi)^n}
\int_0^1 dx \int_0^{1-x} dy \int\frac{d^n k} 
{\left[x k^2+y v_i \cdot k+(1-x-y) v_j \cdot k \right]^3} \, .
\label{I1lp}
\eeq
After several manipulations, Eq. (\ref{I1lp}) becomes
\beqa
I_{1l}&=&\frac{\alpha_s}{\pi} \, (-1)^{-1-\epsilon/2} \, 2^{5\epsilon/2} \, 
\pi^{\epsilon/2} \, \Gamma\left(1+\frac{\epsilon}{2}\right)
(1+\beta^2) \int_0^1 dx \, x^{-1+\epsilon} (1-x)^{-1-\epsilon}
\nonumber \\ && \hspace{-16mm}
{}\times \left\{\int_0^1 dz \left[4z \beta^2 (1-z)+1-\beta^2\right]^{-1}
-\frac{\epsilon}{2} \int_0^1 dz \frac{\ln\left[4z \beta^2 (1-z)
+1-\beta^2\right]}{4z \beta^2 (1-z)+1-\beta^2} 
+{\cal O}\left(\epsilon^2\right)\right\}
\nonumber \\ &&
\label{I1lc}
\eeqa 
where here $\beta=\sqrt{1-4m^2/s}$, with $m$ the quark mass, and 
we have used the relations $v_i \cdot v_j=(1+\beta^2)/2$ 
and $v_i^2=v_j^2=(1-\beta^2)/2$.

The integral over $x$ in Eq. (\ref{I1lc}) contains both ultraviolet (UV) and 
infrared (IR) singularities. We isolate 
the UV singularities and find that 
\beq
\int_0^1 dx \, x^{-1+\epsilon} \, (1-x)^{-1-\epsilon}
=\frac{1}{\epsilon}+{\rm IR} .
\eeq
After calculating the integrals over $z$ in Eq. (\ref{I1lc}), we find 
that the UV poles and constant terms of $I_{1l}$ are 
\beqa
I_{1l}^{UV}&=&\frac{\alpha_s}{\pi} \frac{(1+\beta^2)}{2\beta}
\left\{\frac{1}{\epsilon} \ln\left(\frac{1-\beta}{1+\beta}\right)
+\frac{1}{2} \left(4 \ln 2+\ln \pi-\gamma_E-i\pi\right)
\ln\left(\frac{1-\beta}{1+\beta}\right) \right.
\nonumber \\ && \left.
{}+\frac{1}{4} \ln^2(1+\beta)-\frac{1}{4} \ln^2(1-\beta)
-\frac{1}{2} {\rm Li}_2\left(\frac{1+\beta}{2}\right)
+\frac{1}{2} {\rm Li}_2\left(\frac{1-\beta}{2}\right) \right\}.
\eeqa

Complete one-loop calculations for heavy quark production in axial gauge were 
presented in Ref. \cite{KS}.

\subsection{Two-loop calculation}

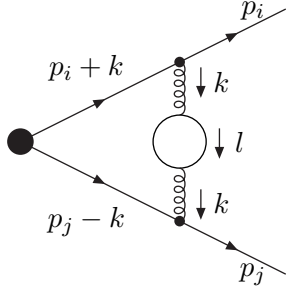
\begin{figure}[htb]
\begin{center}
\begin{picture}(120,120)(0,0)
\Vertex(0,50){5}
\ArrowLine(0,50)(60,80)
\ArrowLine(60,80)(100,100)
\Vertex(60,80){2}  
\Gluon(60,80)(60,60){2}{4}
\Text(25,78)[c]{$p_i+k$}
\Text(88,100)[c]{$p_i$}
\LongArrow(68,78)(68,68)
\Text(76,73)[c]{$k$}
\LongArrow(75,55)(75,45)
\Text(83,50)[c]{$l$}
\GCirc(60,50){10}{1} 
\LongArrow(68,32)(68,22)
\Text(76,27)[c]{$k$}
\ArrowLine(0,50)(60,20)
\ArrowLine(60,20)(100,0)
\Gluon(60,40)(60,20){2}{4}
\Vertex(60,20){2}
\Text(25,20)[c]{$p_j-k$}
\Text(88,0)[c]{$p_j$}
\end{picture}
\end{center}
\caption{\label{quarkl}  Two-loop eikonal diagram, involving a quark loop, 
with outgoing massive quarks.}
\end{figure}

In this subsection we calculate the two-loop integral $I_{2l}$ 
for the quark-loop diagram in Fig.~(\ref{quarkl}) given by
\beqa
I_{2l}&=&(-1) n_f g_s^4 \int\frac{d^n k}{(2\pi)^n}\frac{d^n l}{(2\pi)^n}
\frac{v_i^{\mu}}{v_i\cdot k} \frac{(-v_j^{\rho})}{(-v_j\cdot k)}
\frac{(-i)g_{\mu\nu}}{k^2} \frac{(-i)g_{\rho\sigma}}{k^2} 
\nonumber \\ && \times 
{\rm Tr} \left[-i \gamma^{\nu}
\frac{i l\!\!/}{l^2} (-i) \gamma^{\sigma} i \frac{(l\!\!/-k\!\!/)}
{(l-k)^2}\right] \, .
\label{I2l}
\eeqa
After a few manipulations involving the trace we can write this integral as
\beq 
I_{2l}=-4n_f \frac{g_s^4}{(2\pi)^{2n}} \left[I_{2l}^a+I_{2l}^b+I_{2l}^c
+I_{2l}^d+I_{2l}^e\right]
\label{I2lae}
\eeq
where 
\beq
I_{2l}^a = v_i\cdot v_j \int\frac{d^n k}
{v_i\cdot k \; v_j\cdot k \; k^4} \int \frac{d^n l}{(l-k)^2}
\eeq
\beq
I_{2l}^b = -v_i\cdot v_j \int\frac{d^n k}
{v_i\cdot k \; v_j\cdot k \; k^4} \int d^n l\frac{l\cdot k}{l^2(l-k)^2}
\eeq
\beq
I_{2l}^c = -2 \int\frac{d^n k}{v_i\cdot k \; 
v_j\cdot k \; k^4} \int d^n l\frac{v_i\cdot l \; v_j\cdot l}{l^2(l-k)^2} 
\eeq
\beq
I_{2l}^d = \int\frac{d^n k}{v_i\cdot k  \; k^4} 
\int d^n l\frac{v_i\cdot l}{l^2(l-k)^2}
\eeq
\beq
I_{2l}^e = \int\frac{d^n k}{v_j\cdot k  \; k^4} 
\int d^n l\frac{v_j\cdot l}{l^2(l-k)^2} \, .
\eeq

We begin with the evaluation of $I_{2l}^a$. Since
\beq
\int \frac{d^n l}{(l-k)^2}=0
\label{lzint}
\eeq
we find $I_{2l}^a=0$.

Next we evaluate $I_{2l}^b$. Using Feynman parameterization, we find 
\beq
\int d^n l\frac{l^{\mu}}{l^2(l-k)^2}=i \pi^{(5-\epsilon)/2}\,  2^{-2+\epsilon}
\, \Gamma\left(\frac{\epsilon}{2}\right)\frac{\Gamma\left(1-\frac{\epsilon}{2}
\right)}{\Gamma\left(\frac{3}{2}-\frac{\epsilon}{2}\right)} \, 
(k^2)^{-\epsilon/2} \, k^{\mu}
\label{l1int}
\eeq
therefore
\beq
I_{2l}^b=-i \pi^{(5-\epsilon)/2}\, 2^{-2+\epsilon} \, v_i\cdot v_j \,  
\Gamma\left(\frac{\epsilon}{2}\right)\frac{\Gamma\left(1-\frac{\epsilon}{2}
\right)}{\Gamma\left(\frac{3}{2}-\frac{\epsilon}{2}\right)}
\int\frac{d^n k} {v_i\cdot k \; v_j\cdot k \; (k^2)^{1+\epsilon/2}} \, .
\eeq
The $k$ integral in the above expression is   
\beqa
\int\frac{d^n k} {v_i\cdot k \; v_j\cdot k \; (k^2)^{1+\epsilon/2}}
&=&i \pi^{2-\epsilon/2} \, 2^{2+2\epsilon} \, (-1)^{-1-\epsilon}
\frac{\Gamma(1+\epsilon)}{\Gamma\left(1+\frac{\epsilon}{2}\right)}
\nonumber \\ && \hspace{-35mm}
\times \int_0^1 dx \, x^{-1+2\epsilon} (1-x)^{-1-2\epsilon} 
\int_0^1 dy \left[-2 \beta^2 y^2+2 \beta^2 y
+\frac{1-\beta^2}{2}\right]^{-1-\epsilon}. 
\label{kint}
\eeqa
The integral over $x$ in Eq. (\ref{kint}) contains both UV and IR 
singularities. We isolate the UV singularities and find that 
\beq
\int_0^1 dx \, x^{-1+2\epsilon} (1-x)^{-1-2\epsilon}=\frac{1}{2 \epsilon}+
{\rm IR} \, .
\eeq
The integral over $y$ is given in terms of hypergeometric functions
$_2F_1\left(-\epsilon,1+\epsilon,1-\epsilon,\frac{1\pm\beta}{2}\right)$
which can be expanded in powers of $\epsilon$.
After some calculation, we find 
\beqa
\int\frac{d^n k} {v_i\cdot k \; v_j\cdot k \; (k^2)^{1+\epsilon/2}}
&=&\frac{2 i \pi^2} {\beta \, \epsilon} \ln\left(\frac{1-\beta}{1+\beta}\right)
+\frac{2 i \pi^2} {\beta}
\left[{\rm Li}_2\left(\frac{2}{1+\beta}\right)
-{\rm Li}_2\left(\frac{2}{1-\beta}\right) 
\right. 
\nonumber \\ &&  \hspace{-35mm} \left. 
{}+\ln^2(1+\beta)-\ln^2(1-\beta)+\frac{1}{2}(6 \ln 2-\ln \pi-\gamma_E)
\ln\left(\frac{1-\beta}{1+\beta}\right)\right] .
\label{kintr}
\eeqa
Assembling everything together we find the result for the UV poles 
of $I_{2l}^b$, 
\beqa
I_{2l}^{b \, UV}&=&\pi^4 \frac{(1+\beta^2)}{\beta} \left\{\frac{1}{\epsilon^2} 
\ln\left(\frac{1-\beta}{1+\beta}\right)
+\frac{1}{\epsilon} \left[{\rm Li}_2\left(\frac{2}{1+\beta}\right)
-{\rm Li}_2\left(\frac{2}{1-\beta}\right) 
\right. \right.
\nonumber \\ &&  \hspace{-5mm} \left. \left. 
{}+\ln^2(1+\beta)-\ln^2(1-\beta)+(1+3 \ln 2-\ln \pi-\gamma_E)
\ln\left(\frac{1-\beta}{1+\beta}\right)\right] \right\}.
\eeqa

We continue with the evaluation of $I_{2l}^c$. Now
\beqa
\int d^n l\frac{l^{\mu} l^{\nu}}{l^2(l-k)^2}&=&i \pi^{2-\epsilon/2} \, 
\Gamma\left(\frac{\epsilon}{2}\right) \Gamma\left(1-\frac{\epsilon}{2}\right)
\frac{\Gamma\left(3-\frac{\epsilon}{2}\right)}
{\Gamma(4-\epsilon)} (k^2)^{-\epsilon/2} \, k^{\mu} \, k^{\nu}
\nonumber \\ &&
{}+\frac{i \pi^{2-\epsilon/2}}{2} g^{\mu \nu} 
\Gamma\left(-1+\frac{\epsilon}{2}\right) 
\frac{\left(\Gamma\left(2-\frac{\epsilon}{2}\right)\right)^2}
{\Gamma(4-\epsilon)} (k^2)^{1-\epsilon/2} 
\label{l2int}
\eeqa
and after a few manipulations we find
\beq
I_{2l}^c=-i \pi^{2-\epsilon/2} \Gamma\left(-1+\frac{\epsilon}{2}\right)
\frac{\left(\Gamma\left(2-\frac{\epsilon}{2}\right)\right)^2}
{\Gamma(4-\epsilon)} v_i \cdot v_j
\int\frac{d^n k} {v_i\cdot k \; v_j\cdot k \; (k^2)^{1+\epsilon/2}} \, .
\eeq
The integral over $k$ was evaluated before for $I_{10}^b$, Eq. (\ref{kint}).
We thus find that the UV poles of $I_{10}^c$ are 
\beqa
I_{2l}^{c \, UV}&=&\pi^4 \frac{1+\beta^2}{3\beta}
\left\{-\frac{1}{\epsilon^2}\ln\left(\frac{1-\beta}{1+\beta}\right)
+\frac{1}{\epsilon}\left[-{\rm Li}_2\left(\frac{2}
{1+\beta}\right)+{\rm Li}_2\left(\frac{2}{1-\beta}\right) \right. \right.
\nonumber \\ && \hspace{-15mm} \left. \left.
{}-\ln^2(1+\beta)+\ln^2(1-\beta)
+\left(-\frac{4}{3}-3\ln 2+\ln \pi+\gamma_E \right)
\ln\left(\frac{1-\beta}{1+\beta}\right)\right] \right\}.
\eeqa

Finally, we calculate $I_{2l}^d$ and $I_{2l}^e$. We use Eq. (\ref{l1int}) 
for the $l$ integral and then find that the remaining integral over $k$ 
vanishes, so $I_{2l}^d=I_{2l}^e=0$.

Adding all the terms in Eq. (\ref{I2lae}), the final result for the UV poles 
of $I_{2l}$ is
\beqa
I_{2l}^{UV}&=&-n_f \frac{\alpha_s^2}{\pi^2} \frac{(1+\beta^2)}{6\beta}
\left\{\frac{1}{\epsilon^2} \ln\left(\frac{1-\beta}{1+\beta}\right)
+\frac{1}{\epsilon}\left[{\rm Li}_2\left(\frac{2}{1+\beta}\right)
-{\rm Li}_2\left(\frac{2}{1-\beta}\right) 
\right. \right.
\nonumber \\ &&  \hspace{-10mm} \left. \left.
{}+\ln^2(1+\beta)-\ln^2(1-\beta)
+\left(\frac{5}{6}+5\ln 2+\ln \pi-\gamma_E \right)
\ln\left(\frac{1-\beta}{1+\beta}\right)\right] \right\}.
\label{I2lUV}
\eeqa

More results for two-loop integrals with massive quarks will appear 
in \cite{NKPS}.

\section{Small-$x$ resummations}

In the last forty years there has been a large effort trying to understand 
what are the correct effective degrees of freedom underlying the strong 
interaction at high energies. In scattering processes where the 
center-of-mass energy is much larger than any other scales the  
Balitsky-Fadin-Kuraev-Lipatov (BFKL) 
approach~\cite{BFKL1,BFKL2,BFKL3,BFKL4,BFKL5} emerges as the 
correct approach to describe the scattering. This framework relies upon 
$t$-channel ``Reggeized'' gluons interacting with each other via standard 
gluons in the $s$-channel and a gauge invariant three particle vertex.
This simple structure is a consequence of using multi-Regge kinematics 
where 
gluon cascades are ordered in longitudinal components but with a random walk 
in transverse momenta.  
Although this simple iterative and linear structure must be modified at higher 
energies in order to introduce unitarization and non-linear corrections, 
there is a window at present and future colliders where the BFKL predictions 
hold.

In the leading logarithmic approximation (LLA) we resum terms of the form 
$(\alpha_s \ln{s})^n$. Diagrams contributing to 
the running of the strong coupling do not appear and the coupling is a 
constant parameter. The factor needed to scale the 
energy in the logarithms is also free and the predictability of 
the LL approximation is limited. In the  
next-to-leading logarithmic approximation (NLLA) 
diagrams with an extra power in the
 coupling without introducing an extra logarithm in energy are considered. 
The coupling is allowed to run and the energy scale is determined.

In this contribution we discuss three aspects of the BFKL resummation program. 
In subsection~\ref{coherence} we review the relevant equations to describe 
final states at small values of Bjorken $x$ in Deep Inelastic Scattering 
(DIS). We introduce the concept of color coherence and the CCFM equation. 
We show the differences and similarities between the BFKL approach and the 
introduction of angular ordering in the case of jet rates. 
In subsection~\ref{collinear} we analyse in detail how to extend the region 
of applicability of the multi-Regge kinematics, the basic ingredient in 
the BFKL approach, to regions with collinear emissions. We will find an 
interesting structure in the higher-order corrections that can be resummed 
into a Bessel function of the first kind, which accounts for the double 
logarithms in tranverse scales. In subsection~\ref{conformal} we briefly 
explain the $SL(2,C)$ invariance associated to the BFKL Hamiltonian and 
how it shows up in the physics of multijet events, in particular in 
the production of Mueller-Navelet jets at a hadron collider.

\subsection{QCD coherence and small-$x$ final states}
\label{coherence}

In Quantum Electrodynamics coherence effects are responsible for the suppression of soft bremsstrahlung from 
electron-positron pairs. In QCD processes such as $g \rightarrow q {\bar q}$ any soft gluon emitted with an 
angle from one of the fermionic lines larger than the angle of emission in the $q{\bar q}$ pair will probe 
the total color charge of the pair. This charge is the same as the one from the parent gluon and the radiation 
takes place as if the soft gluon was emitted from it. This color coherence leads to the angular ordering of 
sequential gluon emissions. 

In DIS, let us say that the $(i-1){\rm th}$ emitted gluon from the proton has energy $E_{i-1}$. A gluon 
radiated from it with a fraction $(1-z_{i})$ of its energy and a transverse momentum $q_{i}$ will have an 
opening angle 
\begin{eqnarray}
\theta_{i}\approx \frac{q_{i}}{(1-z_{i})E_{i-1}},
\end{eqnarray}
with
\begin{eqnarray}
z_{i}=\frac{E_{i}}{E_{i-1}}.
\end{eqnarray}
Color coherence leads to angular ordering with increasing opening angles 
towards the hard scale (the photon). Therefore, we have \(\theta_{i+1} 
> \theta_{i}\), or 
\begin{eqnarray}
\frac{q_{i+1}}{1-z_{i+1}}>\frac{z_{i}q_{i}}{1-z_{i}},
\end{eqnarray}
which in the limit $z_{i},z_{i+1}\ll1$ reduces to
\begin{eqnarray}
q_{i+1}>z_{i}q_{i}.
\end{eqnarray}
In Ref. \cite{Catani:1990yc,Marchesini:1995wr,Ciafaloni:1988ur,Catani:1991gu} 
the BFKL equation for the unintegrated structure function was obtained in a form suitable for the study 
of exclusive observables: 
\begin{eqnarray}
f_{\omega}(\mbox{\boldmath $k$})=f_{\omega}^{0}(\mbox{\boldmath $k$})
+\bar\alpha_{S}\int\frac{d^{2}\mbox{\boldmath $q$}}{\pi q^{2}}\int_{0}^{1}
\frac{dz}{z}z^{\omega}\Delta_{R}(z,k)\Theta(q-\mu)f_{\omega}(\mbox{\boldmath 
$q$}+\mbox{\boldmath $k$}),
\end{eqnarray}
where $\mu$ is a collinear cutoff, \mbox{\boldmath $q$} is the transverse 
momentum of the emitted gluon, and the gluon Regge factor is
\begin{eqnarray}
\Delta_{R}(z_{i},k_{i})=\exp\left[-\bar\alpha_{S}\ln\frac{1}{z_{i}}\ln
\frac{k_{i}^{2}}{\mu^2}\right],
\end{eqnarray}
with \(k_i\equiv|\mbox{\boldmath $k$}_{i}|\), and $\bar\alpha_{S}\equiv \alpha_{S} N_c /\pi$. Under iteration, 
this expression generates real gluon emissions with all the virtual corrections summed to all orders. Since 
$f_{\omega}$ is an inclusive structure function, it includes the sum over all final states and the 
$\mu$-dependence cancels between the real and virtual contributions.

The structure function is defined by integrating over all $\mu^{2} \leq q_{i}^{2}\leq Q^{2}$, {\it i.e.}
\begin{eqnarray}
F_{0\omega}(Q,\mu) \equiv \Theta(Q-\mu) + \sum_{r=1}^{\infty}\int_{\mu^{2}}
^{Q^{2}}\prod_{i=1}^{r}\frac{d^{2}\mbox{\boldmath $q$}_{i}}{\pi q_{i}^{2}}
dz_{i}\frac{\bar{\alpha}_{S}}{z_{i}}z_{i}^{\omega}\Delta_{R}(z_{i},k_{i}),
\end{eqnarray}
with $i$ real gluon emissions in each iteration of the kernel. The contributions from a fixed number 
$r$ of emitted gluons is 
\begin{eqnarray}
F_{0\omega}(Q) = \int_{0}^{1} dx ~x^{\omega} F_{0}(x,Q) = 1 + 
\sum_{r=1}^{\infty} F_{0\omega}^{(r)}(Q).
\end{eqnarray}
In Ref.~\cite{Marchesini:1995wr} the perturbative expansion for the $F_{0\omega}^{(r)}(Q,\mu)$
\begin{eqnarray}
F^{(r)}_{0\omega}(Q,\mu)=\sum_{n=r}^{\infty}C^{(r)}_{0}(n;T)
\frac{\bar\alpha_{S}^{n}}{\omega^{n}},
\end{eqnarray}
was obtained with $T \equiv \ln({Q/\mu})$. Then we have
\begin{eqnarray}
F_{0\omega}(Q)\equiv\sum_{i=0}^{\infty}F_{0 \omega}^{(i)}(Q)=
\left(\frac{Q^2}{\mu^{2}}\right)^{\bar\gamma},
\end{eqnarray}
where $\bar\gamma$ is the BFKL anomalous dimension. It was pointed out that 
coherence effects significantly modify the individual $F_{0\omega}^{(r)}(Q)$ whilst preserving the sum 
$F_{0\omega}(Q)$, and care must be taken to account properly for coherence in the 
calculation of associated distributions. 

Modifying the BFKL formalism to account for 
coherence~\cite{Catani:1990yc,Marchesini:1995wr,Ciafaloni:1988ur,Catani:1991gu}, $F_{0\omega}(Q,\mu)$ becomes
\begin{eqnarray}
F_{\omega}(Q,\mu) = \Theta(Q-\mu) + \sum_{r=1}^{\infty}\int_{0}^{Q^{2}}
\prod_{i=1}^{r}\frac{d^{2}\mbox{\boldmath $q$}_{i}}{\pi q_{i}^{2}}dz_{i}
\frac{\bar{\alpha}_{S}}{z_{i}}z_{i}^{\omega}\Delta(z_{i},q_{i},k_{i})
\Theta(q_{i}-z_{i-1}q_{i-1}),
\end{eqnarray}
where $\Delta_{R}(z_{i},k_{i})$ is substituted by
\begin{eqnarray}
\Delta(z_{i},q_{i},k_{i})=\exp\left[-\bar\alpha_{S}\ln\frac{1}{z_{i}}\ln
\frac{k_{i}^{2}}{z_{i}q_{i}^{2}}\right];~ ~ k_{i} > q_{i},
\label{constraint}
\end{eqnarray}
and for the first emission we take $q_{0}z_{0} = \mu$. The expansion of $F_{\omega}^{(r)}(Q)$ is now 
\begin{eqnarray}
F_{\omega}^{(r)}(Q)=\sum_{n=r}^{\infty}\sum_{m=1}^{n}C^{(r)}(n,m;T)\frac{
\bar\alpha_{S}^{n}}{\omega^{2n-m}}.
\end{eqnarray}
A collinear cutoff is needed only on the emission of the first gluon because subsequent collinear 
emissions are regulated by the angular ordering constraint. 

The rates for the emission of a fixed number of resolved gluons, with a transverse momentum larger 
than a resolution scale $\mu_{R}$, together with any number 
of unresolved ones, were calculated in Ref.~\cite{Forshaw:1998uq} 
in the LLA, to third order in $\bar\alpha_S$.
$\mu_{R}$ is constrained by the collinear cutoff and the hard scale, $\mu \ll \mu_{R}\ll Q$. For the $n$-jet 
rate all the graphs with $n$ resolved gluons and any number of unresolvable ones were considered.
Expanding the Regge factors to ${\cal O} ({\bar \alpha}_{S}^{3})$ we find that the jet rates both in the 
multi-Regge (BFKL) approach and in the coherent (CCFM) approach are the same:
\begin{eqnarray}
{\rm 0 ~jet} &=& \frac{(2\bar{\alpha}_{S})}{\omega}S
+\frac{(2\bar{\alpha}_{S})^{2}}{\omega^{2}}\left[\frac{S^{2}}{2}\right]
+\frac{(2\bar{\alpha}_{S})^{3}}{\omega^{3}}\left[\frac{S^{3}}{6}\right], \\  
{\rm 1~jet} &=& \frac{(2\bar{\alpha}_{S})}{\omega}T+\frac{(2\bar{\alpha}
_{S})^{2}}
{\omega^{2}}\left[TS-\frac{1}{2}T^{2}\right]+\frac{(2\bar{\alpha}_{S})^{3}}
{\omega^{3}}\left[\frac{1}{3}T^{3}-\frac{1}{2}T^{2}S+\frac{1}{2}TS^{2}\right],\\
{\rm 2~jet} &=& \frac{(2\bar{\alpha}_{S})^{2}}{\omega^{2}}
\left[T^{2}\right]+\frac{(2\bar{\alpha}_{S})^{3}}{\omega^{3}}
\left[T^{2}S-\frac{7}{6}T^{3}\right],\\
{\rm 3~jet} &=& \frac{(2\bar{\alpha}_{S})^{3}}{\omega^{3}}\left[T^{3}\right],
\end{eqnarray}
with $T=\ln(Q/\mu_R)$ and $S=\ln(\mu_R/\mu)$. When coherence is introduced the singularities at 
$\omega \sim 0$ are stronger than in the BFKL approach but the extra logarithms cancel in the sum of 
all the graphs needed for the jet rates. The net effect is that the final results are the same as those 
obtained without coherence~\cite{Forshaw:1998uq}. This is true to all orders in the 
coupling~\cite{Forshaw:1999yh} since a generating function for the jet multiplicity 
distribution was obtained in~\cite{Webber:1998we}. Within the multi-Regge kinematics the $r$-jet rate reads
\begin{eqnarray}
R^{(n ~{\rm jet})}_{\omega}(Q,\mu_R) = \frac{F^{(n {\rm ~jet})}_{\omega}(Q,\mu_R,\mu)}{F_{\omega}(Q,\mu)} = 
\frac{1}{n!} \left. \frac{\partial^n}{\partial u^n} R_{\omega}(u,T)\right|_{u=0},
\label{easy}
\end{eqnarray}
where the jet-rate generating function $R_{\omega}$ is given by
\begin{eqnarray}
R_{\omega}(u,T) = \exp{\left(-\frac{2 {\bar \alpha}_s}{\omega} T \right)} 
\left[1+(1-u)\frac{2 {\bar \alpha}_s}{\omega}T\right]^{\frac{u}{1-u}}.
\end{eqnarray}
The same generating function is obtained when coherence is considered. The mean number
of jets and the mean square fluctuation in this number are
\begin{eqnarray}
\langle n\rangle =
\left.\frac{\partial}{\partial u}R_{\omega}(u,T)\right|_{u=1} =
\frac{2 {\bar \alpha}_s}{\omega}T +\frac{1}{2} \left(\frac{2{\bar \alpha}_s}{\omega}T\right)^2 ,
\end{eqnarray}
\begin{eqnarray}
\langle n^2\rangle -\langle n\rangle^2 = \frac{2{\bar \alpha}_s}{\omega}T
+\frac{3}{2} \left(\frac{2{\bar \alpha}_s}{\omega}T\right)^2
+\frac{2}{3}\left(\frac{2{\bar \alpha}_s}{\omega}T\right)^3.
\end{eqnarray}
In general, the $p$th central moment of the jet multiplicity
distribution is a polynomial in ${\bar \alpha}_s T/\omega$ of degree $2p-1$,
indicating that the distribution becomes relatively narrow in
the limit of very small $x$ and large $Q/\mu_{R}$~\cite{Webber:1998we}.

In Ref.~\cite{Ewerz:1999fn,Ewerz:1999tt} the subject was developed even further and all subleading logarithms 
of $Q^{2}/\mu^{2}_{R}$ were included to calculate the jet multiplicity in Higgs production at the LHC. In 
Ref.~\cite{Ewerz:1999fn} they extended the results from a $[{\bar \alpha}_S \ln(1/x) \ln(Q^{2}/\mu_{R}^2)]^n$ resummation 
to a $[{\bar \alpha}_s \ln(1/x)]^n[\ln(Q^{2}/\mu_{R}^2)]^m$ one with $0 < m \leq n$, proving that the quadratic and cubic 
forms of the mean and the variance remain valid. It has also been shown that for any sufficiently inclusive 
observables the CCFM formalism leads to the same results as the BFKL equation~\cite{Salam:1999ft}. The key idea 
to understand this result comes if we try to obtain the results of Ref. \cite{Marchesini:1995wr} from those 
in the previous section in the limit $\mu_{R} \rightarrow 0$. To get the right solution we should consider 
subleading terms ${\bar \alpha}_s \ln^{2}(Q/\mu_R)$ which must be resummed when taking the limit $\mu_{R} \rightarrow 0$ 
to obtain a continuous transition from the case where BFKL and CCFM results are equivalent, to that of them 
being different. If we also consider the effects of introducing the $z \rightarrow 1$ divergent part of the 
splitting function in the CCFM approach we will see that this leads to all BFKL and CCFM final-state properties 
being identical in the $[{\bar \alpha}_s \ln(1/x) \ln(Q^{2}/\mu_{R}^2)]^n$ approximation \cite{Salam:1999ft}. Recent 
reviews devoted to the implementation of CCFM in Monte Carlo event generators can be found in, {\it e.g.},~\cite{Dittmar:2005ed,Alekhin:2005dx,Alekhin:2005dy,Andersen:2006pg}. An 
approach which has the potential to apply BFKL in the NLLA to DIS 
phenomenology is that in 
Ref.~\cite{Andersen:2003an,Andersen:2003wy,Andersen:2004uj,Andersen:2004tt}. 
In the NLLA approximation it is important to carefully take into account 
$k_t$ factorization~\cite{Bartels:2006hg}.

\subsection{Improving the collinear region of multi-Regge kinematics}
\label{collinear}

In this section we revisit the approach of Ref.~\cite{Salam:1998tj} where the multi-Regge kinematics was 
extended to include collinear contributions present to all orders in the BFKL formalism. 
In Ref.~\cite{Vera:2005jt} the structure in transverse momentum space of the double logarithms resummed 
was explicitly extracted. A new renormalization group (RG)-improved kernel was obtained which does not mix transverse with 
longitudinal momentum components. 

In the ${\overline{\rm MS}}$ renormalisation scheme, the BFKL kernel in the NLLA acting on a smooth 
function~\cite{Fadin:1998py,Ciafaloni:1998gs} is
\begin{eqnarray}
\label{ktKernel}
&&\hspace{-1.2cm}\int d^2 \vec{q}_2 \, {\cal K}\left(\vec{q}_1,\vec{q}_2\right) 
f\left({q}_2^2\right) = \nonumber\\
&&\hspace{-1cm}\int \frac{d^2 \vec{q}_2}{\left|{q}_1^2-{q}_2^2\right|} \left\{\left[{\bar \alpha}_s+{\bar \alpha}_s^2\left({\cal S}-\frac{\beta_0}{4 N_c}\ln{\left(
\frac{\left|{q}_1^2-{q}_2^2\right|^2}
{{\rm max}\left({q}_1^2,{q}_2^2\right)\mu^2}\right)}
\right)\right]\right.\nonumber\\
&&\hspace{-1cm}\times\left.\left(f\left({q}_2^2\right)-2 \frac{{\rm min}\left({q}_1^2,{q}_2^2\right)}{\left({q}_1^2+{q}_2^2\right)}f\left({q}_1^2\right)\right)-\frac{{\bar \alpha}_s^2}{4}\left({\cal T}\left({q}_1^2,{q}_2^2\right)+\ln^2{\left(\frac{{q}_1^2}{{q}_2^2}\right)}\right)f\left({q}_2^2\right)\right\}, 
\end{eqnarray}
where $\beta_0 = \left(11 N_c - 2 n_f\right)/3$,
${\cal S} = \left(4-\pi^2 +5 \beta_0/N_c \right)/12$, and ${\cal T} ({q}_1^2,{q}_2^2)$ can be found in 
Ref.~\cite{Fadin:1998py}. The collinear structure can be obtained acting on 
the eigenfunctions in the LLA, {\it i.e.}
\begin{eqnarray}
\label{nondiag}
\int d^2 \vec{q}_2 \, {\cal K} \left(\vec{q}_1,\vec{q}_2\right)
\left(\frac{{\bar \alpha}_s \left({q}_2^2\right)}{{\bar \alpha}_s \left({q}_1^2\right)}\right)^{-\frac{1}{2}}
\left(\frac{{q}_2^2}{{q}_1^2}\right)^{\gamma-1} = 
{\bar \alpha}_s \left({q}_1^2\right) \chi_0\left(\gamma\right)
+ {\bar \alpha}_s^2 \chi_1 \left(\gamma\right).
\end{eqnarray}
Here we have
\begin{eqnarray}
\label{LOeigen}
\chi_0 \left(\gamma\right) &=& 2 \psi(1) - \psi \left(\gamma\right)-
\psi \left(1-\gamma\right), 
\end{eqnarray}
\begin{eqnarray}
\label{chi1}
\chi_1 \left(\gamma\right) &=& {\cal S} \chi_0 \left(\gamma\right)
+ \frac{1}{4}\left(\psi''\left(\gamma\right)+ \psi''\left(1-\gamma\right)\right)
- \frac{1}{4}\left(\phi\left(\gamma\right)+ \phi\left(1-\gamma\right)\right)\\
&&\hspace{-2cm}-\frac{\pi^2 \cos{(\pi \gamma)}}{4 \sin^2(\pi \gamma)(1-2 \gamma)}
\left(3+\left(1+ \frac{n_f}{N_c^3}\right)\frac{(2+3\gamma(1-\gamma))}{(3-2\gamma)(1+2\gamma)}\right) 
+ \frac{3}{2} \zeta_3 - \frac{\beta_0}{8 N_c} \chi_0^2 \left(\gamma\right),  \nonumber
\end{eqnarray}
with $\psi \left(\gamma \right) = \Gamma'\left(\gamma\right)/ 
\Gamma \left(\gamma\right)$ and
\begin{eqnarray}
\phi \left(\gamma\right) + \phi \left(1-\gamma\right) &=&\nonumber\\ 
&&\hspace{-2cm}\sum_{m=0}^{\infty} \left(\frac{1}{\gamma+m}+\frac{1}{1-\gamma+m}\right)
\left(\psi'\left(\frac{2+m}{2}\right)-\psi'\left(\frac{1+m}{2}\right)\right).
\end{eqnarray}
The pole structure around $\gamma = 0,1$ is 
\begin{eqnarray}
\chi_0 \left(\gamma\right) &\simeq& \frac{1}{\gamma} +
\left\{\gamma \rightarrow 1- \gamma\right\},\\
\chi_1 \left(\gamma \right) &\simeq& 
\frac{\rm a}{\gamma}+\frac{\rm b}{\gamma^2}-\frac{1}{2 \gamma^3} +
\left\{\gamma \rightarrow 1- \gamma\right\}
\end{eqnarray}
with 
\begin{eqnarray}
\label{ab}
{\rm a} &=& \frac{5}{12}\frac{\beta_0}{N_c} -\frac{13}{36}\frac{n_f}{N_c^3}
-\frac{55}{36}, \, \, \,
{\rm b} ~=~ -\frac{1}{8}\frac{\beta_0}{N_c} -\frac{n_f}{6 N_c^3}
-\frac{11}{12}.
\end{eqnarray}
The cubic poles stem from $\psi''$ and compensate for the equivalent terms appearing when the Regge-like energy scale 
$s_0 = q_1 q_2$ is shifted to the DIS choice $s_0 = q_{1,2}^2$. Higher 
order terms beyond the NLLA, not compatible with RG evolution, 
are also generated by this change of scale. The NLLA truncation of the 
perturbative expansion is then the reason why the gluon Green's function 
develops oscillations, where the Green's function can have negative values, in the $q_1^2/q_2^2$ ratio.

It is possible to remove the most dominant 
poles in $\gamma$-space incompatible with RG evolution by simply shifting 
the $\omega$-pole present in the BFKL scale invariant 
eigenfunction. Here we focus on the scheme proposed in Ref.~\cite{Salam:1998tj}:
\begin{eqnarray}
\omega \hspace{-2mm}&=& \hspace{-2mm}{\bar \alpha}_s \left(1+\left({\rm a}+\frac{\pi^2}{6}\right){\bar \alpha}_s\right) \left(2 \psi(1)-\psi\left(\gamma+\frac{\omega}{2}-{\rm b}\,{\bar \alpha}_s \right)-\psi\left(1-\gamma+\frac{\omega}{2}-{\rm b}\,{\bar \alpha}_s \right)\right)\nonumber\\&&
{}+ {\bar \alpha}_s^2 \left(\chi_1 \left(\gamma\right) 
+\left(\frac{1}{2}\chi_0\left(\gamma\right)-{\rm b}\right)\left(\psi'(\gamma)+\psi'(1-\gamma)\right)-\left({\rm a}+\frac{\pi^2}{6}\right)\chi_0(\gamma)\right).
\end{eqnarray}
We can approximately solve this equation considering the 
$\omega$-shift in the form
\begin{eqnarray}
\omega = {\bar \alpha}_s  \left(1+ {\rm A} \,{\bar \alpha}_s \right)
\left(2 \psi(1)-\psi\left(\gamma+\frac{\omega}{2}+ {\rm B} \,{\bar \alpha}_s\right)-\psi\left(1-\gamma+\frac{\omega}{2}
+ {\rm B} \,{\bar \alpha}_s\right)\right),
\end{eqnarray}
which can be written as
\beq
\label{generalshift}
\omega={\bar \alpha}_s  \left(1+ {\rm A} {\bar \alpha}_s\right)
\sum_{m=0}^{\infty} \left(\frac{1}{\gamma + m+\frac{\omega}{2}+ {\rm B} \,{\bar \alpha}_s}+
\frac{1}{1-\gamma+m+\frac{\omega}{2}+ {\rm B} \,{\bar \alpha}_s}-\frac{2}{m+1}\right).
\eeq
The solution to this shift can be obtained by adding all the 
approximated solutions at the different poles plus a term related 
to the virtual contributions, {\it i.e.}
\begin{eqnarray}
\label{assumption}
\omega &=& \sum_{m=0}^{\infty} 
\left\{-(1+ 2 m + 2 \,{\rm B} \,{\bar \alpha}_s)+
\left|\gamma + m + {\rm B} \,{\bar \alpha}_s\right|
\left(1+\frac{2{\bar \alpha}_s \left(1+ {\rm A} {\bar \alpha}_s\right)}{\left(\gamma + m + {\rm B} \,{\bar \alpha}_s\right)^2}\right)^{\frac{1}{2}}\right. \nonumber\\
&&\hspace{-1cm}\left.+
\left|1-\gamma + m + {\rm B} \,{\bar \alpha}_s\right|
\left(1+\frac{2{\bar \alpha}_s \left(1+ {\rm A} {\bar \alpha}_s\right)}{\left(1-\gamma + m + {\rm B} \,{\bar \alpha}_s\right)^2}\right)^{\frac{1}{2}}
- \frac{2 {\bar \alpha}_s \left(1+ {\rm A} {\bar \alpha}_s\right)}{m+1}\right\}.
\end{eqnarray}
At the $\gamma =0,1$ poles this expansion generates the NLLA terms:
\begin{eqnarray}
\omega \simeq \frac{{\bar \alpha}_s}{\gamma} + {\bar \alpha}_s^2 \left(\frac{\rm A}{\gamma}
- \frac{\rm B}{\gamma^2}- \frac{1}{2 \gamma^3} \right)+ \left\{\gamma \rightarrow 1-\gamma\right\}. 
\end{eqnarray}
To match the original kernel at NLLA we set 
${\rm A} = {\rm a}$ and ${\rm B} = - {\rm b}$ from Eq.~(\ref{ab}).

We now include the full NLLA scale invariant kernel without double counting terms:
\begin{eqnarray}
\label{All-poles}
\omega &=& \bar{\alpha}_s \chi_0 (\gamma) + \bar{\alpha}_s^2 \chi_1 (\gamma) \\
&+& \left\{\sum_{m=0}^{\infty} \left[\left(\sum_{n=0}^{\infty}
\frac{(-1)^n (2n)!}{2^n n! (n+1)!}\frac{\left({\bar \alpha}_s+ {\rm a} \,{\bar \alpha}_s^2\right)^{n+1}}{\left(\gamma + m - {\rm b} \,{\bar \alpha}_s\right)^{2n+1}}\right) \right. \right. \nonumber\\&&\left.\left.-\frac{\bar{\alpha}_s}{\gamma + m} - \bar{\alpha}_s^2 \left(\frac{\rm a}{\gamma +m} + \frac{\rm b}{(\gamma + m)^2}-\frac{1}{2(\gamma+m)^3}\right)\right]+ \left\{\gamma \rightarrow 1-\gamma\right\}\right\}. \nonumber
\end{eqnarray}
\begin{figure}
\centering
\includegraphics[width=0.35\textwidth,angle=-90]{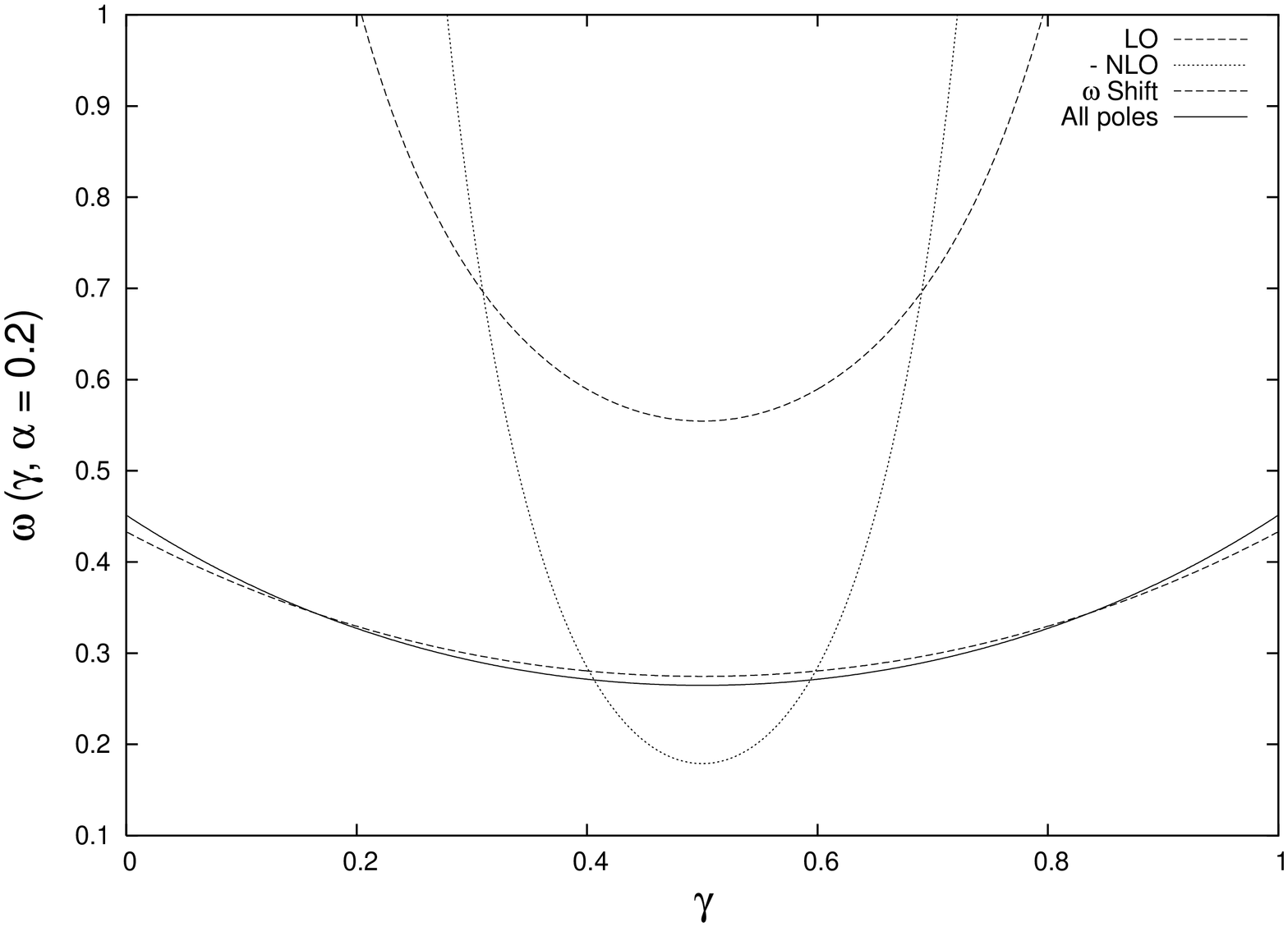}\includegraphics[width=0.35\textwidth,angle=-90]{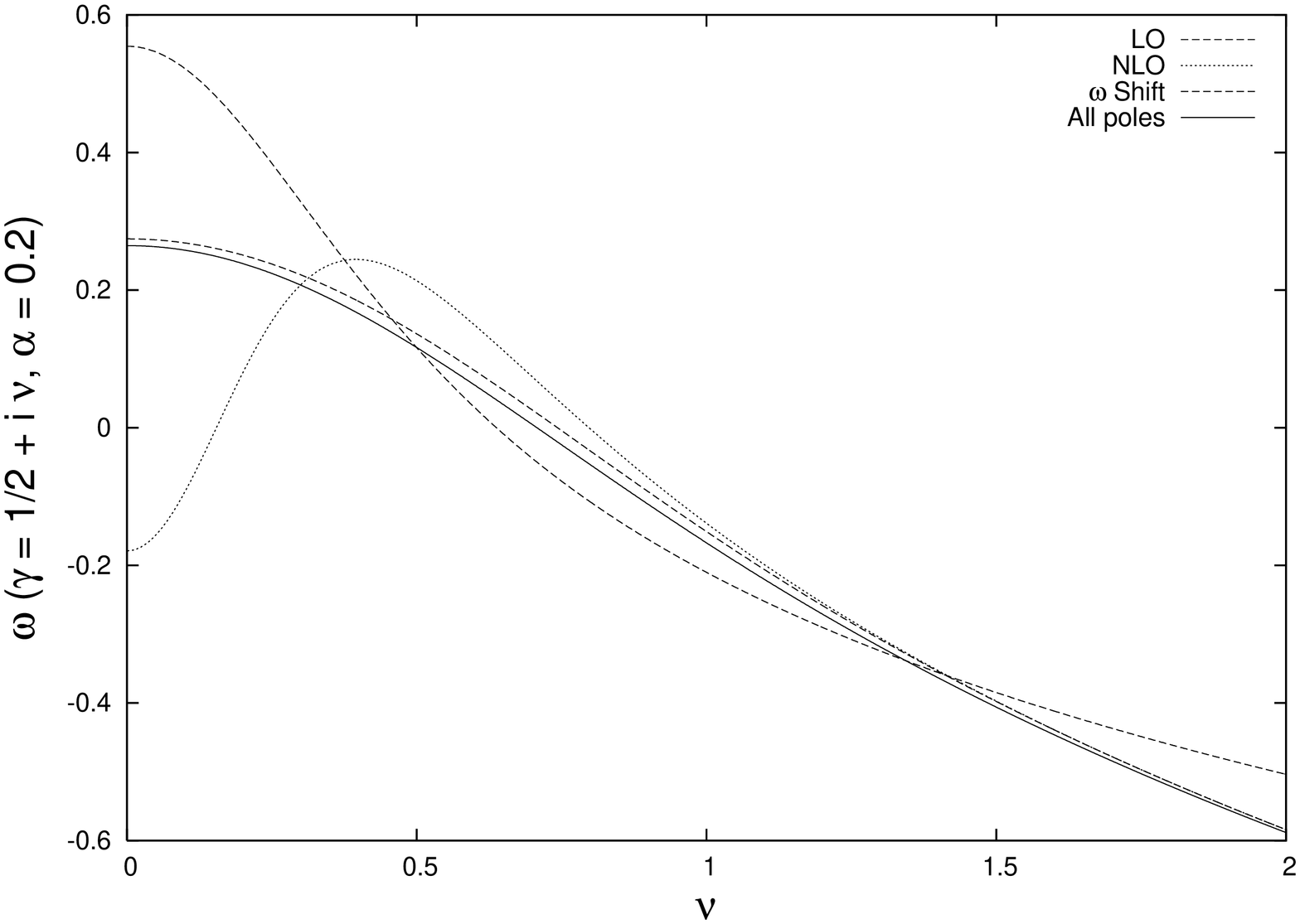}
\caption[]{$\gamma$-representation of the LLA and NLLA kernels. The RG--improved kernel by a $\omega$-shift 
is included together with the new ``all-poles'' approximation.}
\label{FullFigure1}
\end{figure}
This result reproduces the $\omega$-shift very closely, 
see Fig.~\ref{FullFigure1}. The imaginary part of $\gamma$ at the 
maximum of the NLLA scale invariant eigenvalue 
(middle plot of Fig.~\ref{FullFigure1}) is not 
zero and results in oscillations in the $q_1^2/q_2^2$ variable. These are 
eliminated when the RG--improved kernel is used, as it 
also happens for the ``all--poles'' kernel. 

It is very important to note that in Eq.~(\ref{All-poles}) the 
$\omega$-space is decoupled from the $\gamma$-representation. In 
Ref.~\cite{Vera:2005jt} an expression for the collinearly improved BFKL kernel 
which does not mix longitudinal with transverse degrees of freedom was found. 
The only modification needed in the full NLLA kernel to introduce the 
``all-poles'' resummation is to remove the term
\begin{eqnarray}
\label{presc1}
-\frac{\bar{\alpha}_s^2}{4}\frac{1}{(\vec{q}-\vec{k})^2}
\ln^2\left({\frac{q^2}{k^2}}\right)
\end{eqnarray}
 in the real emission kernel, ${\cal K}_r \left(\vec{q},\vec{k}\right)$, 
and replace it with
\begin{eqnarray}
\label{presc2}
\frac{1}{(\vec{q}-\vec{k})^2} \left\{\left(\frac{q^2}{k^2}\right)^{-{\rm b}{\bar \alpha}_s 
\frac{\left|k-q\right|}{k-q}}
\sqrt{\frac{2\left({\bar \alpha}_s+ {\rm a} \,{\bar \alpha}_s^2\right)}{\ln^2{\left(\frac{q^2}{k^2}\right)}}} 
J_1 \left(\sqrt{2\left({\bar \alpha}_s+ {\rm a} \,{\bar \alpha}_s^2\right) 
\ln^{2}{\left(\frac{q^2}{k^2}\right)}}\right) \right.\nonumber\\
&& \left.\hspace{-7cm}- {\bar \alpha}_s - {\rm a} \, {\bar \alpha}_s^2
+ {\rm b} \, {\bar \alpha}_s^2 \frac{\left|k-q\right|}{k-q}
\ln{\left(\frac{q^2}{k^2}\right)} \right\},
\end{eqnarray}
with $J_1$ the Bessel function of the first kind.
When the difference between the $q^2$ and $k^2$ scales is not very large then 
\begin{eqnarray}
J_1 \left(\sqrt{2 {\bar \alpha}_s 
\ln^{2}{\left(\frac{q^2}{k^2}\right)}}\right) &\simeq& 
\sqrt{\frac{{\bar \alpha}_s}{2} \ln^{2}{\left(\frac{q^2}{k^2}\right)}},
\end{eqnarray}
and its influence is minimal, not affecting the ``Regge--like'' region. When 
the ratio of transverse momenta becomes larger then 
\begin{eqnarray}
J_1 \left(\sqrt{2 {\bar \alpha}_s  
\ln^{2}{\left(\frac{q^2}{k^2}\right)}}\right) \simeq 
\left(\frac{2}{\pi^2 {\bar \alpha}_s  
\ln^{2}{\left(\frac{q^2}{k^2}\right)}}\right)^{\frac{1}{4}}
\cos{\left(\sqrt{2 {\bar \alpha}_s  
\ln^{2}{\left(\frac{q^2}{k^2}\right)}}-\frac{3\pi}{4}\right)}
\end{eqnarray}
compensating for the unphysical oscillations. This resummation of all-poles 
has been applied to extend the region of applicability of BFKL calculations in 
the NLLA in the case of 
electroproduction of light vector mesons in Ref.~\cite{Caporale:2007vs}.

\subsection{Conformal signatures at the Large Hadron Collider: azimuthal angle}
\label{conformal}

We now proceed to review the work of Ref.~\cite{Vera:2006un} where azimuthal 
angle decorrelations in inclusive dijet cross sections were studied 
analytically to include the NLLA to the BFKL kernel, while keeping the jet 
vertices at leading order. It was shown how the angular decorrelation for 
jets with a wide relative separation in rapidity largely decreases when higher 
order effects are considered.

Observables where BFKL effects should be dominant require a large enough 
center-of-mass energy, and two large and similar transverse scales. An 
example is the inclusive hadroproduction of two jets with  large and similar 
transverse momenta  and a large relative separation in rapidity, Y, the 
so-called Mueller-Navelet jets, first proposed in 
Ref.~\cite{Mueller:1986ey}. A rise with Y in the partonic cross section 
was predicted in agreement with the LLA hard Pomeron intercept. At the 
hadronic level, Mueller-Navelet jets are produced in a region where the 
parton distribution falls very quickly, reducing this rise. 
Small $x$ resummation effects are very relevant if we investigate the 
azimuthal angle decorrelation of the pair of jets. BFKL enhances soft real 
emission as Y increases, reducing the angular correlation. This was first 
investigated in the LLA in 
Ref.~\cite{DelDuca:1993mn,DelDuca:1994ng,Stirling:1994zs}. The rate of 
decorrelation in the LLA lies quite below the experimental 
data~\cite{Abachi:1996et,Abbott:1999ai,Abbott:1997nf,Abe:1996mj} at the 
Tevatron. This motivates the NLLA discussion of this subsection.

We are interested in the cross section parton + parton $\rightarrow$ jet + 
jet + soft emission, with the two jets having transverse momenta $\vec{q}_1$ 
and $\vec{q}_2$ and with a relative rapidity separation Y. The differential 
partonic cross section is
\begin{eqnarray}
\frac{d {\hat \sigma}}{d^2\vec{q}_1 d^2\vec{q}_2} &=& \frac{\pi^2 {\bar \alpha}_s^2}{2} 
\frac{f \left(\vec{q}_1,\vec{q}_2,{\rm Y}\right)}{q_1^2 q_2^2}.
\end{eqnarray}
It is useful to introduce a Mellin transform:
\begin{eqnarray}
f \left(\vec{q}_1,\vec{q}_2,{\rm Y}\right) &=& \int \frac{d\omega}{2 \pi i} e^{\omega {\rm Y}} f_\omega \left(\vec{q}_1,\vec{q}_2\right).
\end{eqnarray}
The solution to the BFKL equation in the LLA is
\begin{eqnarray}
f_\omega \left(\vec{q}_1,\vec{q}_2\right) &=& \frac{1}{2 \pi^2} \sum_{n = -\infty}^\infty
\int_{-\infty}^\infty d \nu \,{\left(q_1^2\right)}^{-i \nu -\frac{1}{2}} {\left(q_2^2\right)}^{i \nu -\frac{1}{2}} \frac{e^{i n \left(\theta_1-\theta_2\right)}}{\omega - {\bar \alpha}_s \chi_0 \left(\left|n\right|,\nu\right)} 
\end{eqnarray}
with
\begin{eqnarray}
\chi_0 \left(n,\nu\right) &=& 2 \psi \left(1\right) - \psi \left(\frac{1}{2}+ i \nu + \frac{n}{2}\right) - \psi\left(\frac{1}{2}- i \nu +\frac{n}{2}\right). 
\end{eqnarray}
The nonforward BFKL equation corresponds to a 
Schr{\"o}dinger-like equation with a holomorphically separable Hamiltonian 
where $- i \,{\rm Y}$ is the time variable. Both the holomorphic and 
antiholomorphic sectors are invariant under spin zero M{\"o}bius 
transformations with eigenfunctions carrying a conformal weight of the form 
$\gamma = \frac{1}{2} + i \nu + \frac{n}{2} $. In the principal series of the 
unitary representation, $\nu$ is real and $\left|n\right|$ the integer 
conformal spin~\cite{Lipatov:1985uk}. In this way extracting information 
about $n$ is equivalent to proving the conformal structure of high energy QCD.
 
We now integrate over the phase space of the two emitted gluons together with 
some general jet vertices, {\it i.e.}
\begin{eqnarray}
{\hat \sigma} \left(\alpha_s, {\rm Y},p^2_{1,2}\right) &=&
\int d^2{\vec{q}_1} \int d^2{\vec{q}_2} \,
\Phi_{\rm jet_1}\left(\vec{q}_1,p_1^2\right)
\,\Phi_{\rm jet_2}\left(\vec{q}_2,p_2^2\right)\frac{d {\hat \sigma}}{d^2\vec{q}_1 d^2\vec{q}_2}.
\end{eqnarray}
In the jet vertices only leading-order terms are kept:
\begin{eqnarray}
\Phi_{\rm jet_i}^{(0)} \left(\vec{q},p_i^2\right)&=& \theta \left(q^2-p_i^2\right), \label{eqn:jetLO}
\end{eqnarray}
where $p_i^2$ corresponds to a resolution scale for the transverse momentum of 
the gluon jet. To extend this analysis it is needed to use the NLO jet 
vertices in Ref.~\cite{Bartels:2001ge,Bartels:2002yj} where 
the definition of a jet is much more complex than Eq.~(\ref{eqn:jetLO}). We 
then have
\begin{eqnarray}
{\hat \sigma} \left(\alpha_s, {\rm Y},p_{1,2}^2\right) =
\frac{\pi^2 {\bar \alpha}_s^2}{2} \int d^2{\vec{q}_1} \int d^2{\vec{q}_2} \,
\frac{\Phi_{\rm jet_1}^{(0)}\left(\vec{q}_1,p_1^2\right)}{q_1^2}
\,\frac{\Phi_{\rm jet_2}^{(0)}\left(\vec{q}_2,p_2^2\right)}{q_2^2}
f \left(\vec{q}_1,\vec{q}_2,{\rm Y}\right). 
\end{eqnarray}
In a transverse momenta operator representation:
\begin{eqnarray}
\left< \vec{q}\right|\left.\nu,n\right> &=& \frac{1}{\pi \sqrt{2}} 
\left(q^2\right)^{i \nu -\frac{1}{2}} \, e^{i n \theta}, 
\label{eignfns}
\end{eqnarray}
the action of the NLO kernel, calculated in Ref.~\cite{Kotikov:2000pm}, is 
\begin{eqnarray}
{\hat K} \left|\nu,n\right> &=& \left\{\frac{}{}{\bar \alpha}_s \, \chi_0\left(\left|n\right|,\nu\right) + {\bar \alpha}_s^2 \, \chi_1\left(\left|n\right|,\nu\right) \right.\nonumber\\
&&\left.\hspace{-2cm}+\,{\bar \alpha}_s^2 \,\frac{\beta_0}{8 N_c}\left[2\,\chi_0\left(\left|n\right|,\nu\right) \left(i \frac{\partial}{\partial \nu}+ \log{\mu^2}\right)+\left(i\frac{\partial}{\partial \nu}\chi_0\left(\left|n\right|,\nu\right)\right)\right]\right\} \left|\nu,n\right>,
\label{opKernel}
\end{eqnarray}
where $\chi_1$, for a general conformal spin, reads 
\begin{eqnarray}
\chi_1\left(n,\gamma \right) &=& {\cal S} \chi_0 \left(n, \gamma\right)
+ \frac{3}{2} \zeta\left(3\right) - \frac{\beta_0}{8 N_c}\chi_0^2\left(n,\gamma\right)\nonumber\\
&+&\frac{1}{4}\left[\psi''\left(\gamma+\frac{n}{2}\right)+\psi''\left(1-\gamma+\frac{n}{2}\right)-2 \,\phi\left(n,\gamma\right)-2 \,\phi\left(n,1-\gamma\right)\right]\nonumber\\
&-&\frac{\pi^2 \cos{\left(\pi \gamma\right)}}{4 \sin^2\left(\pi \gamma\right)\left(1-2\gamma\right)}\left\{\left[3+\left(1+\frac{n_f}{N_c^3}\right)\frac{2+3\gamma\left(1-\gamma\right)}{\left(3-2\gamma\right)\left(1+2\gamma\right)}\right]\delta_{n 0}\right.\nonumber\\
&&\left.\hspace{2cm}-\left(1+\frac{n_f}{N_c^3}\right)\frac{\gamma\left(1-\gamma\right)}{2\left(3-2\gamma\right)\left(1+2\gamma\right)}\delta_{n 2}\right\}.
\end{eqnarray}
The function $\phi$ can be found in Ref.~\cite{Kotikov:2000pm}.

\begin{figure}
\centering
\includegraphics[width=0.35\textwidth,angle=-90]{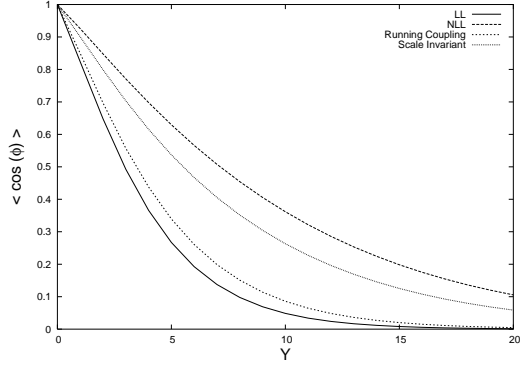}
\caption[]{Dijet azimuthal angle decorrelation as a function of their separation in rapidity.}
\label{Cos1Y}
\end{figure}

The jet vertices on the basis in Eq.~(\ref{eignfns}) are:
\begin{eqnarray}
\int d^2{\vec{q}}\,\frac{\Phi_{\rm jet_1}^{(0)}\left(\vec{q},p_1^2\right)}{q^2} \left<\vec{q}\right.\left|\nu,n\right> = 
\frac{1}{\sqrt{2}}\frac{1}{\left(\frac{1}{2}-i \nu\right)}\left(p_1^2\right)^{i \nu - \frac{1}{2}} \delta_{n,0} \equiv c_1\left(\nu\right) \delta_{n,0},
\label{IFproj}
\end{eqnarray}
with the $c_2\left(\nu\right)$ projection of $\Phi_{\rm jet_2}^{(0)}$ on 
$\left<n,\nu\right|\left.\vec{q}\right>$ being the complex conjugate 
of~(\ref{IFproj}) with $p_1^2$ being replaced by $p_2^2$. The cross section 
can then be rewritten as
\begin{eqnarray}
{\hat \sigma} \left(\alpha_s, {\rm Y},p_{1,2}^2\right) &=&
\frac{\pi^2 {\bar \alpha}_s^2}{2} \sum_{n=-\infty}^\infty \int_{-\infty}^\infty d \nu 
\,e^{{\bar \alpha}_s \chi_0\left(\left|n\right|,\nu\right) {\rm Y}} 
c_1\left(\nu\right) c_2\left(\nu\right) \delta_{n,0} \label{logdercross}\\
&&\hspace{-3.3cm}\times \left\{1+{\bar \alpha}_s^2 \, {\rm Y} \left[\chi_1\left(\left|n\right|,\nu\right)+\frac{\beta_0}{4 N_c} \left(\log{(\mu^2)}+ \frac{i}{2} \frac{\partial}{\partial \nu}\log{\left(\frac{c_1\left(\nu\right)}{c_2\left(\nu\right)}\right)}+ \frac{i}{2} \frac{\partial}{\partial \nu}\right)\chi_0\left(\left|n\right|,\nu\right)\right]\right\}.\nonumber
\end{eqnarray}
For the LO jet vertices the logarithmic derivative in Eq.~(\ref{logdercross}) 
is 
\begin{eqnarray}
- i \frac{\partial}{\partial \nu}\log{\left(\frac{c_1\left(\nu\right)}{c_2\left(\nu\right)}\right)} &=& \log{\left(p_1^2p_2^2\right)}+ \frac{1}{\frac{1}{4}+\nu^2}.
\end{eqnarray}
If $\phi = \theta_1-\theta_2 - \pi$, in the case of two equal resolution 
momenta, $p_1^2 = p_2^2 \equiv p^2$, the angular differential cross section 
can be expressed as
\begin{eqnarray}
\frac{d {\hat \sigma}\left(\alpha_s, {\rm Y},p^2\right)}{d \phi}  &=&
\frac{\pi^3 {\bar \alpha}_s^2}{2 p^2} \frac{1}{2 \pi}\sum_{n=-\infty}^\infty 
e^{i n \phi} {\cal C}_n \left({\rm Y}\right),
\end{eqnarray}
with
\begin{eqnarray}
{\cal C}_n \left({\rm Y}\right) =
\int_{-\infty}^\infty \frac{d \nu}{2 \pi}\frac{e^{{\bar \alpha}_s \left(p^2\right){\rm Y} \left(\chi_0\left(\left|n\right|,\nu\right)+{\bar \alpha}_s  \left(p^2\right) \left(\chi_1\left(\left|n\right|,\nu\right)-\frac{\beta_0}{8 N_c} \frac{\chi_0\left(\left|n\right|,\nu\right)}{\left(\frac{1}{4}+\nu^2\right)}\right)\right)}}{\left(\frac{1}{4}+\nu^2\right)}.
\label{Cn}
\end{eqnarray}
The coefficient governing the energy dependence of the cross section 
corresponds to $n=0$:
\begin{eqnarray}
{\hat \sigma}\left(\alpha_s, {\rm Y},p^2\right) &=& 
\frac{\pi^3 {\bar \alpha}_s^2}{2 p^2} \, {\cal C}_0 \left({\rm Y}\right).
\end{eqnarray}
We have chosen the resolution scale $p = 30 \, {\rm GeV}$, $n_f = 4$ and 
$\Lambda_{\rm QCD} = 0.1416$ GeV. The $n = 0$ coefficient is directly related 
to the normalized cross section
\begin{eqnarray}
\frac{{\hat \sigma} \left({\rm Y}\right)}{{\hat \sigma} \left(0\right)} 
&=& \frac{{\cal C}_0 \left({\rm Y}\right)}{{\cal C}_0\left(0\right)}. 
\end{eqnarray}
\begin{figure}
\centering
\includegraphics[width=0.35\textwidth,angle=-90]{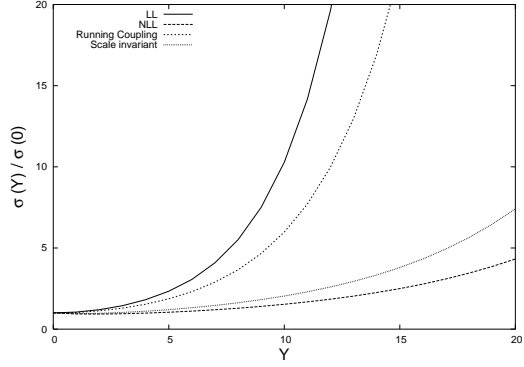}
\caption[]{Evolution of the partonic cross section with the rapidity separation of the dijets.}
\label{SectionconY}
\end{figure}
The rise with Y of this observable is shown in 
Fig.~\ref{SectionconY}. Clearly the NLL intercept is very much reduced with 
respect to the LL case. The remaining coefficients with $n \geq 1$ all 
decrease with Y. Because of this, the angular correlations also 
diminish as the rapidity interval between the jets gets larger. This point 
can be studied in detail using the mean values 
\begin{eqnarray}
\left<\cos{\left( m \phi \right)}\right> &=& \frac{{\cal C}_m \left({\rm Y}\right)}{{\cal C}_0\left({\rm Y}\right)}.
\end{eqnarray}
$\left<\cos{\left(\phi\right)}\right>$ is calculated in Fig.~\ref{Cos1Y}. The 
NLL effects decrease the azimuthal angle decorrelation. This is the case for 
the running of the coupling and also for the scale invariant terms. This is 
encouraging from the phenomenological point of view given that the data at the 
Tevatron typically have lower decorrelation than predicted by LLA BFKL or LLA 
with running coupling. 
The difference in the decorrelation between LLA and NLLA is driven by the $n=0$ 
conformal spin since the ratio
\begin{eqnarray}
\frac{\left<\cos{\left(\phi \right)}\right>^{\rm NLLA}}{\left<\cos{\left(\phi \right)}\right>^{\rm LLA}} &=& \frac{{\cal C}_1^{\rm NLLA} \left({\rm Y}\right)}{{\cal C}_0^{\rm NLLA}\left({\rm Y}\right)}\frac{{\cal C}_0^{\rm LLA} \left({\rm Y}\right)}{{\cal C}_1^{\rm LLA}\left({\rm Y}\right)},
\end{eqnarray}
remains in the region
\begin{eqnarray}
1.2 > \frac{{\cal C}_1^{\rm NLLA} \left({\rm Y}\right)}{{\cal C}_1^{\rm LLA}\left({\rm Y}\right)} > 1.
\end{eqnarray}
\begin{figure}
\centering
\includegraphics[width=0.35\textwidth,angle=-90]{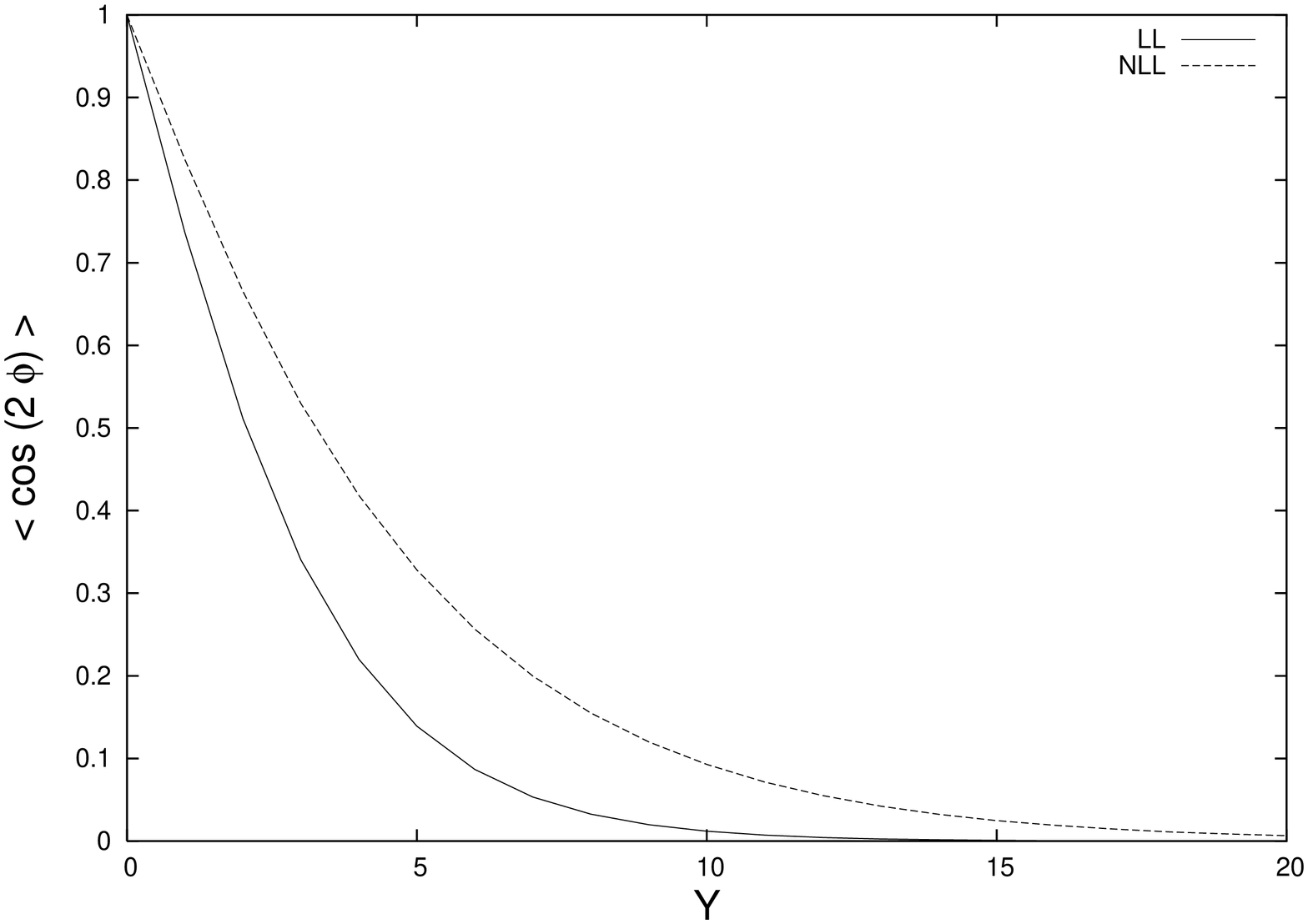}\includegraphics[width=0.35\textwidth,angle=-90]{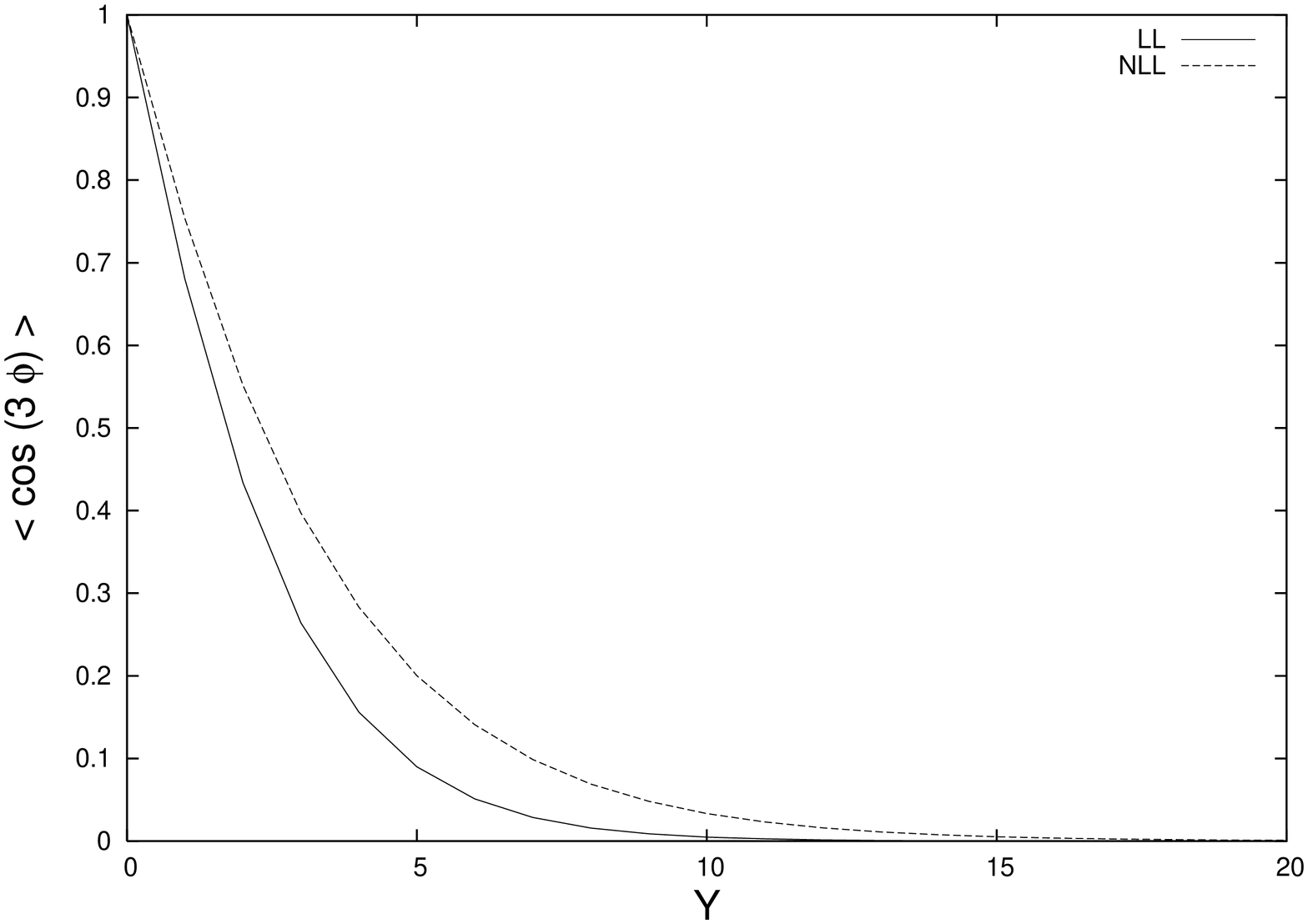}
\caption[]{Dijet azimuthal angle decorrelation as a function of their separation in rapidity.}
\label{Cos23Y}
\end{figure}
This is a consequence of the good convergence, in terms of asymptotic 
intercepts of the NLLA BFKL calculation, for conformal spins larger than zero. 
For completeness the $m=2,3$ cases for 
$\left<\cos{\left(m \phi\right)}\right>$ are shown in Fig.~\ref{Cos23Y}. 
These distributions test the structure of the 
higher conformal spins. The methods of this subsection have been applied to 
phenomenology of dijets at the Tevatron and the LHC 
in~\cite{Vera:2007kn,Marquet:2007xx}, and to the production of forward jets 
in DIS at HERA in~\cite{Vera:2007dr}.

\section{Conclusion}

The precision of perturbative QCD calculations will play a major role
in the confidence  of new physics discoveries, both at this generation of
experiments, Tevatron and LHC, and in future experiments. The most available 
avenue of improving the precision of QCD is through resummation of large
contributions. We have presented results for the resummation of large-$x$
contributions and separately small-$x$ contributions. In both cases, the large
contributions arise from incomplete cancellations of virtual and real terms,  
and can be computed in the eikonal approximation.

We have shown that the inclusion of soft-gluon corrections to top quark 
production cross sections is essential to stabilize the unphysical scale 
variations in the order-by-order calculations. 
This is necessary for any sort of precision
calculation of the top mass and production channels. Additionally, we have 
shown the importance of resummation on $W$ production at large transverse 
momentum, and on Higgs production. Discovery of the Higgs boson is the last
remaining test of the Standard Model and precision measurements of its 
properties is essential to proceed forward with beyond the Standard Model
theories.

We have also presented a framework to include collinear effects into the 
BFKL formalism. This stabilizes the oscillatory behavior that arises when
one moves away from the strict kinematic regime of validity. It was shown how 
this inclusion improves the prediction of Mueller-Navalet jets, jets
with a large rapidity separation but similar transverse scales. This is a 
process which will be observed at the LHC where the BFKL formalism should
flourish; an important test of the complex behavior of QCD. A comparison 
between the predictions steming from a pure BFKL analysis and one including 
QCD coherence in multijet final states in DIS has been also discussed in 
detail.

\section*{Acknowledgements}
  
The work of N.K. was supported by the National Science Foundation under
Grant No. PHY~0555372.

\end{document}